\documentclass[lettersize,journal]{IEEEtran}
\usepackage{amsmath, amsfonts, amssymb}
\usepackage{amsthm}   
\usepackage{algpseudocode}
\usepackage{algorithmicx,algorithm}
\usepackage{array}
\usepackage[caption=false,font=normalsize,labelfont=sf,textfont=sf]{subfig}
\usepackage{textcomp}
\usepackage{stfloats}
\usepackage{url}
\usepackage{verbatim}
\usepackage{graphicx}
\usepackage{cite}
\usepackage{bm}
\usepackage{tabularx}
\usepackage{threeparttable}
\usepackage{mathtools}

\newcommand{\eref}[1]{(\ref{#1})}

\newcommand{\cref}[1]{Constraint~\ref{#1}}

\newcommand{\tref}[1]{TABLE~\ref{#1}}

\newtheorem{remark}{Remark}

\usepackage{tikz} 
\usepackage[utf8]{inputenc}
\usepackage{pgfplots} 
\makeatletter
\newcommand{\gettikzxy}[3]{%
  \tikz@scan@one@point\pgfutil@firstofone#1\relax
  \edef#2{\the\pgf@x}%
  \edef#3{\the\pgf@y}%
}

\begin{document}

\title{LEO Constellations as a Decentralized GNSS Network: \\ Optimizing PNT Corrections in Space}

\author{Xing Liu,
Xue Xian Zheng,
José A. López-Salcedo,~\IEEEmembership{Senior Member,~IEEE,}
Tareq Y. Al-Naffouri,~\IEEEmembership{Fellow,~IEEE,}
Gonzalo Seco-Granados,~\IEEEmembership{Fellow,~IEEE}
\thanks{Xing Liu, José A. López-Salcedo, and Gonzalo Seco-Granados are with the Department of Telecommunications and Systems Engineering, Universitat Autònoma de Barcelona (UAB), and also with the Institute for Space Studies of Catalonia (IEEC), Barcelona, Spain.}

\thanks{Xue Xian Zheng and Tareq Y. Al-Naffouri are with the Electrical and Computer Engineering Program, Computer, Electrical and Mathematical Sciences and Engineering Division, King Abdullah University of Science and Technology (KAUST), Thuwal 23955-6900, Saudi Arabia.}

\thanks{This work has been partly supported by the Catalan Government in the framework of the Catalonia Space 2030 Strategy and the AGAUR--ICREA Academia Program, and by the Spanish Agency of Research (AEI) under Grant PID2023-152820OB-I00 funded by MICIU/AEI/10.13039/501100011033 and by ERDF/EU. This work has also been partly supported by the KAUST Office of Sponsored Research (OSR) under Award No. RFS-CRG12-2024-6478.}

\thanks{Corresponding author: Xue Xian Zheng (xuexian.zheng@kaust.edu.sa).}
}

\markboth{Journal of \LaTeX\ Class Files,~Vol.~x, No.~x, October~2026}%
{Shell \MakeLowercase{\textit{et al.}}: A Sample Article Using IEEEtran.cls for IEEE Journals}


\maketitle

\begin{abstract}
With the rapid expansion of low Earth orbit (LEO) constellations, thousands of satellites are now in operation, many equipped with onboard GNSS receivers capable of continuous orbit determination and time synchronization. This development is creating an unprecedented spaceborne GNSS network, offering new opportunities for network-driven precise LEO orbit and clock estimation. Yet, current onboard GNSS processing is largely standalone and often insufficient for high-precision applications, while centralized fusion is challenging due to computational bottlenecks and the lack of in-orbit infrastructure. In this work, we report a decentralized GNSS network over large-scale LEO constellations, where each satellite processes its own measurements while exchanging compact information with neighboring nodes to enable precise orbit and time determination. We model the moving constellation as a dynamic graph and employ a momentum-accelerated gradient tracking (GT) method to achieve stable convergence despite topology changes. Numerical simulations with constellations containing hundreds of satellites show that the proposed method matches the accuracy of an ideal centralized benchmark. Ultimately, this framework supports the development of autonomous and self-organizing space systems, enabling high-precision navigation with reduced dependence on continuous ground contact. 
\end{abstract}


\section{Introduction}
\IEEEPARstart{L}{ow} Earth orbit (LEO) satellites have become a transformative platform for communications, navigation, remote sensing, Earth observation, and numerous other space applications~\cite{9681631,9110855,9502642,reid2025xona,10945753,9984725}. Many of these applications require accurate knowledge of the satellite’s position and clock state, and onboard Global Navigation Satellite System (GNSS) receivers have become a widely used solution for orbit determination and time synchronization~\cite{selvan2023precise,allahvirdi2022precise,4319208,montenbruck2025gnss}. Currently, most LEO-onboard GNSS processing still relies on standalone positioning using broadcast ephemerides embedded within GNSS signals, which provides only meter-level accuracy due to the limited precision of the broadcast products. Missions requiring higher accuracy instead depend on precise GNSS products generated by International GNSS Service (IGS) analysis centers and derived from terrestrial continuously operating reference station (CORS) networks, which provide decimeter to centimeter accuracy~\cite{selvan2023precise,montenbruck2005rapid}. However, such performance is typically achieved only in post-processed orbit determination, since most LEO spacecraft cannot maintain continuous ground connectivity and the latency of precise products limits their usability for real-time operations~\cite{selvan2023precise,allahvirdi2022precise}. Consequently, broadcast ephemerides offer real-time availability but limited positioning accuracy, whereas precise products deliver substantially higher accuracy at the cost of latency and intermittent accessibility. An increasing number of LEO missions, including location-enabled communication systems, formation flying, positioning, navigation, and timing (PNT) services, satellite altimetry, and spaceborne sensing, now require real-time precise orbit and clock determination~\cite{selvan2023precise}. Bridging this accuracy-latency gap requires a new architecture capable of delivering both high precision and real-time availability.

\begin{figure*}[ht]
  \centering
 \includegraphics[width=\textwidth]{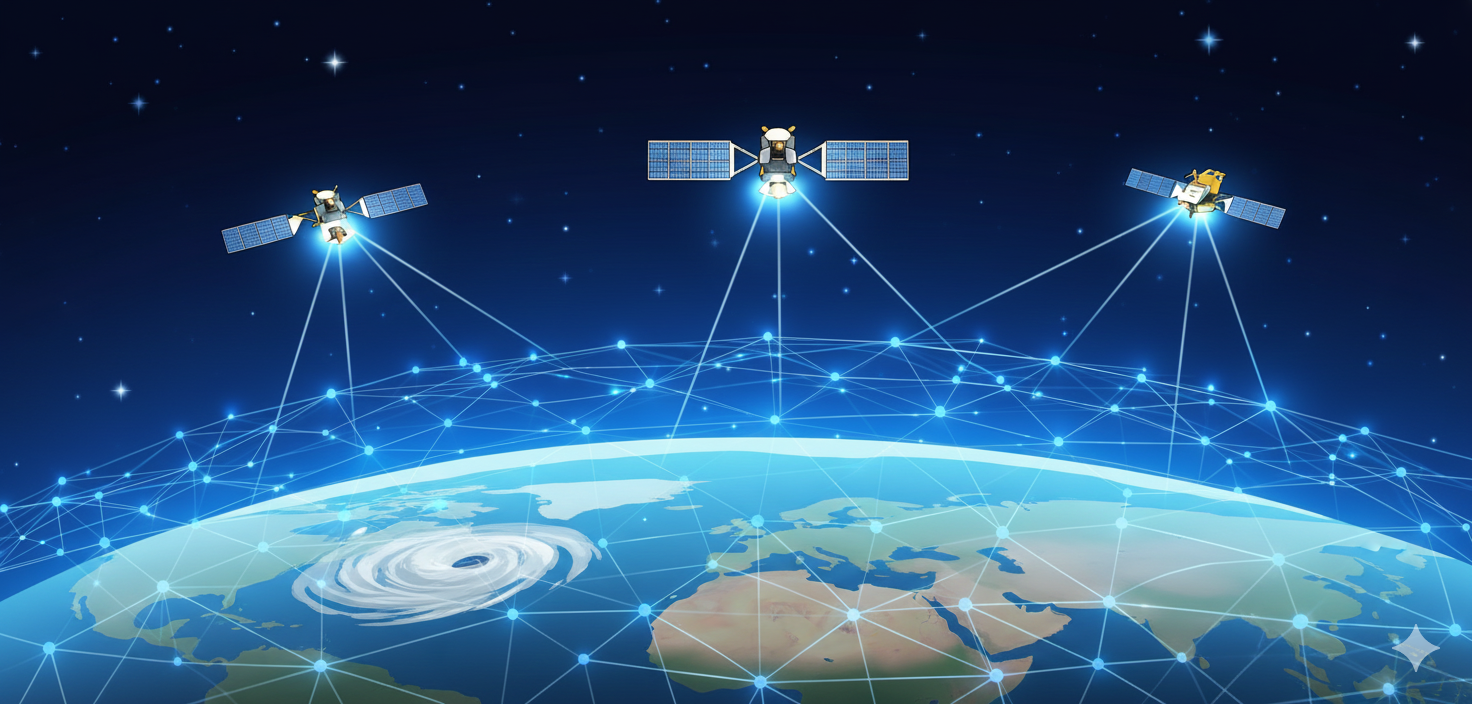} 
  \caption{Conceptual illustration of a LEO constellation viewed as a decentralized spaceborne GNSS network. LEO satellites (bright nodes) form a sparse, time-varying inter-satellite communication graph, while each LEO node receives signals from visible GNSS satellites.}
  \label{fig:leo_network_concept}
\end{figure*}

Ground-based GNSS networks demonstrate the effectiveness of cooperative processing: observations from global CORS networks are jointly processed at centralized analysis centers to generate precise satellite clocks and orbits, atmospheric corrections, and bias models, which underpin high-precision positioning techniques such as precise point positioning (PPP)~\cite{heroux1995gps,zumberge1997precise} and network-assisted PPP real-time kinematic (PPP-RTK)~\cite{wabbena2005ppp}. Modern LEO constellations equipped with onboard GNSS receivers can, in principle, be viewed as a spaceborne analogue of such networks, in which hundreds or thousands of rapidly moving satellites collectively provide a global and geometry-rich set of GNSS observations, as conceptually illustrated in Fig.~\ref{fig:leo_network_concept}. However, existing LEO onboard GNSS receivers typically operate independently: GNSS-derived information is not exchanged among satellites, orbit and clock states are not estimated at the constellation level, and cooperative onboard processing across the constellation remains largely unexplored~\cite{selvan2023precise}. Inspired by the success of terrestrial CORS networks, this work examines whether a LEO constellation can operate cooperatively to jointly estimate its orbit and clock states and to generate network-derived GNSS correction information directly onboard. Such a system could extend network-based GNSS processing to the space segment and enable the onboard generation of network-derived GNSS satellite clock corrections, which may serve as a supporting component for high-precision positioning applications, particularly in remote or otherwise challenging environments.

LEO provides a particularly favorable environment for spaceborne GNSS-based inference. Operating above the dense layers of the atmosphere, LEO satellites are effectively free from tropospheric delay, experience substantially reduced ionospheric effects due to shorter signal propagation paths, and are largely unaffected by multipath~\cite{hofmann2008gnss,allahvirdi2022precise}. As a result, GNSS measurements collected in LEO are considerably cleaner than those obtained by ground-based receivers. In addition, the high orbital velocity of LEO spacecraft induces strong Doppler signatures and rapidly varying observation geometry, both of which enhance measurement observability and render Doppler a highly informative and complementary observable~\cite{palomo2019space}. The scale of modern LEO constellations further amplifies these advantages, as thousands of rapidly moving satellites naturally provide dense global coverage and rich geometric diversity. 

Despite these favorable observation conditions, realizing cooperative orbit and clock estimation at the constellation level is fundamentally challenged by the dynamic nature of LEO networks. Satellite visibility and inter-satellite connectivity vary over time, while scaling to thousands of satellites dramatically increases the density of observations, placing stringent demands on update rates~\cite{9838251,11020624}. These characteristics give rise to a large-scale, communication-constrained network with limited bandwidth and time-varying connectivity. Under these constraints, centralized processing is challenging: it incurs excessive communication and computational burdens, introduces significant latency, and ultimately fails to scale with constellation size. Moreover, no such spaceborne fusion center currently exists, and establishing one for constellation-wide processing would require substantial cost and dedicated infrastructure. Together, these challenges motivate a decentralized processing paradigm, in which each LEO satellite performs local estimation based on its own observations and exchanges only compact information with neighboring nodes over a time-varying communication graph.

With sufficient connectivity, decentralized estimation offers a pathway to globally consistent solutions without forwarding raw measurements to a centralized analysis center. By shifting computation toward the data sources, such approaches reduce backhaul load, lower latency, and enable inherent parallelism with more efficient resource utilization~\cite{sayed2014diffusion,4118472,6994854,mcmahan2017communication}. Beyond these benefits, decentralized architectures alleviate communication and computational burdens, improve scalability, enhance resilience to node or link failures, preserve data locality, and support higher update rates as observation density increases~\cite{yang2019survey}. These properties make decentralized processing a promising direction for large, dynamic LEO constellations operating as cooperative GNSS observation networks. Recent work has advanced decentralized estimation of satellite correction products using ground-receiver networks~\cite{zheng2025decentralized}. Nevertheless, a principled treatment of system-wide coordination under decentralized architectures remains lacking in large-scale, communication-constrained LEO networks.

Based on these considerations, this work investigates the idea of treating a LEO constellation as a decentralized spaceborne GNSS network, in which onboard GNSS receivers cooperatively estimate constellation-level states to enable precise LEO orbit and clock determination. In this formulation, each satellite operates as a node in a time-varying communication graph and contributes its locally collected GNSS observations to a constellation-wide estimation process. This perspective naturally leads to a decentralized estimation framework designed to accommodate the large scale and high dynamics of LEO constellations, as well as the intermittent connectivity inherent to onboard operations.

The main contributions of this work are as follows:
\begin{enumerate}
\item We propose a spaceborne GNSS network architecture for large-scale LEO constellations, in which LEO satellites cooperatively process shared GNSS information to perform precise self-orbit and clock determination in a fully decentralized manner.
\item We establish the constellation-level observation model, analyze its estimability, and construct a common estimable parameterization with clearly defined local and shared variables, yielding a well-defined distributed optimization formulation for orbit and clock determination of LEO satellites.
\item We model the inter-satellite communication topology induced by LEO orbital dynamics as a time-varying but predictable graph, and adopt a momentum-accelerated gradient tracking (GT) algorithm to solve the resulting distributed estimation problem through neighbor-to-neighbor information exchange.
\item We provide a large-scale numerical evaluation with hundreds of LEO satellites, demonstrating that decentralized precise orbit and clock determination can achieve accuracy comparable to centralized processing. These findings suggest a viable path toward spaceborne PNT correction generation, where scalability and time-varying inter-satellite connectivity are key considerations.
\end{enumerate}

The remainder of this paper is organized as follows. Section~\ref{sec:observation} introduces the spaceborne GNSS observation model. Section~\ref{sec:estimability} analyzes the estimability of parameters in LEO onboard GNSS networks. Section~\ref{sec:formulation_algorithm} presents the decentralized estimation formulation and its solution using gradient tracking over time-varying inter-satellite communication graphs. Section~\ref{sec:numerical_results} reports numerical results that demonstrate the feasibility of the proposed approach. Section~\ref{sec:conclusion} concludes the paper.

\section{Spaceborne GNSS Observation Equations}
\label{sec:observation}
LEO satellites, whose orbital altitudes typically range from 300 to 1500 km, experience high relative dynamics with GNSS satellites, producing Doppler shifts of up to tens of kilohertz. These strong Doppler dynamics provide rich information content and, when properly modeled, improve the observability of orbit, velocity, and onboard clock states, in contrast to ground receivers, which observe only weak Doppler variations. In addition, LEO satellites operate above the troposphere, so no tropospheric delay is present, and their much shorter ionospheric path greatly reduces ionospheric delay. Together, the high dynamics and simplified propagation conditions create an observation environment that differs from that of ground-based GNSS.

In the proposed spaceborne GNSS networked framework, a large LEO constellation is modeled as a network of spaceborne GNSS receiver nodes. Each LEO satellite is equipped with an onboard GNSS receiver that tracks signals transmitted by GNSS satellites. Consider a system consisting of $L$ LEO satellites and $G$ GNSS satellites. Due to orbital geometry and line-of-sight (LOS) visibility, each receiver has access to only a subset of the GNSS constellation at a given epoch. When receiver $l$ observes transmitter $g$, the linearized observation equations on frequency $f$ can be written in unified form as:
\begin{equation}
\begin{aligned}
\mathrm{E} \left\{ \Delta \phi_{l,f}^g  \right\} 
= & \left(\mathbf{u}_l^g\right)^{\mathrm{T}} \Delta \mathbf{p}_l 
+ \Delta t_{l} 
+ \lambda_f \delta_{l,f} - \Delta t^{g} \\
&                               
- \lambda_f \delta_{f}^{g} 
- \mu_f I_{l}^{g} 
+ \lambda_f z_{l,f}^{g}, \\
\mathrm{E} \left\{ \Delta \rho_{l,f}^g  \right\} 
= & \left(\mathbf{u}_l^g\right)^{\mathrm{T}} \Delta \mathbf{p}_l
+ \Delta t_{l} 
+ b_{l,f} - \Delta t^{g} \\
&                                  
- b_f^{g} 
+ \mu_f I_{l}^{g}, \\
\mathrm{E} \left\{\Delta d_{l,f}^g  \right\} 
= & -\frac{1}{\lambda_f} \left(\mathbf{u}_l^g\right)^{\top}\Delta \mathbf{v}_l -\frac{1}{\lambda_f}\left(\Delta \dot{t}_l-\Delta \dot{t}^g\right)\\
& -\frac{\mu_f}{\lambda_f} \dot{I}_l^g,
\end{aligned}
\end{equation}
where $\mathrm{E}\{\cdot\}$ denotes expectation, and
\begin{itemize}
    \item $\Delta \phi_{l,f}^g$, $\Delta \rho_{l,f}^g$, and $\Delta d_{l,f}^g$ denote the undifferenced observed-minus-computed carrier-phase, pseudorange (code), and Doppler measurements, respectively.
    
    \item $l = 1,\dots,L$, $g = 1,\dots,G$, and $f = 1,\dots,F$ index the LEO receivers, GNSS satellites, and frequencies.
    
    \item $\mathbf{u}_{l}^{g} \in \mathbb{R}^3$ denotes the line-of-sight (LOS) unit vector from GNSS satellite $g$ to receiver $l$.
    
    \item $\Delta \mathbf{p}_l \in \mathbb{R}^3$ and $\Delta \mathbf{v}_l \in \mathbb{R}^3$ represent the incremental position and velocity of receiver $l$, respectively.
    
    \item $\Delta t_l$ and $\Delta t^g$ denote the receiver and GNSS satellite clock biases, with $\Delta \dot{t}_l$ and $\Delta \dot{t}^g$ representing the corresponding clock drift terms.
    
    \item $\delta_{l,f}$ and $\delta_f^g$ denote the receiver and GNSS satellite carrier-phase hardware biases at frequency $f$, while $b_{l,f}$ and $b_f^g$ denote the corresponding pseudorange hardware biases.
    
    \item $I_l^g$ represents the ionospheric delay at the first frequency, scaled to frequency $f$ by $\mu_f = \lambda_f^2/\lambda_1^2$. The ionospheric delay rate $\dot{I}_l^g$ is neglected in the subsequent analysis.
    
    \item $z_{l,f}^g$ denotes the carrier-phase ambiguity.
\end{itemize}

To construct the network-wide observation vector, the observables are ordered as follows. 
For each receiver $l$, the measurements are grouped by frequency: starting from 
$f=1$, all carrier-phase observations are listed first, followed by all pseudorange 
observations, and then all Doppler observations. Within each observation type, satellites 
are arranged according to the prescribed satellite ordering. The same structure is repeated 
for frequencies $2$ through $F$, and then for receivers $1$ through $L$. With this data 
organization, the network-wide observation vector is defined as
\begin{equation}
\label{eq:y_net}
\mathbf{y} =\left[\mathbf{y}_1^\mathrm{T}, \cdots, \mathbf{y}_L^\mathrm{T} \right]^\mathrm{T} \in  \mathbb{R}^{m},
\end{equation}
where $m=3F\sum_{l=1}^{L}G_l$ is the total number of observations. The vector $\mathbf{y}_l$ collects the measurements associated with receiver $l$ and is structured as
\[
\mathbf{y}_l
=
\left[
\mathbf{y}_{l,1}^\mathrm{T},
\cdots,
\mathbf{y}_{l,F}^\mathrm{T}
\right]^\mathrm{T} \in \mathbb{R}^{3FG_l},
\]
\[
\mathbf{y}_{l,f}
=
\left[
\pmb{\Delta}\pmb{\phi}_{l,f}^\mathrm{T},
\pmb{\Delta}\pmb{\rho}_{l,f}^\mathrm{T},
\pmb{\Delta}\mathbf{d}_{l,f}^\mathrm{T}
\right]^\mathrm{T} \in \mathbb{R}^{3G_l}.
\]
Here, $\pmb{\Delta}\pmb{\phi}_{l,f},\; \pmb{\Delta}\pmb{\rho}_{l,f},\; \pmb{\Delta}\mathbf{d}_{l,f} \in \mathbb{R}^{G_l}$ contain the frequency-$f$ observations from the $G_l$ GNSS satellites visible to receiver~$l$ (with $G_l < G$).

The unknown parameters of the considered spaceborne GNSS system can be grouped into five physically meaningful categories:
\begin{itemize}
\item \textbf{LEO orbit–related states:}
incremental position and velocity corrections of the LEO satellites.
\[
\mathbf{x}_{\mathrm{leo\text{-}orb}}
=
\operatorname{col}\big\{
\Delta\mathbf{p}_1,\;
\Delta\mathbf{v}_1,\;
\ldots,\;
\Delta\mathbf{p}_L,\;
\Delta\mathbf{v}_L
\big\}
\in\mathbb{R}^{6L},
\]
where $\operatorname{col}\{\cdot\}$ denotes the column-wise concatenation.

\item \textbf{LEO clock and hardware–related states:}
onboard receiver clock offsets and drifts, and carrier-phase and pseudorange hardware biases for each frequency.
\[
\mathbf{x}_{\mathrm{leo\text{-}clk}}
=
\operatorname{col}\big\{
\mathbf{x}_{\mathrm{leo\text{-}clk}}^{(1)},\;
\ldots,\;
\mathbf{x}_{\mathrm{leo\text{-}clk}}^{(L)}
\big\}
\in\mathbb{R}^{L(2+2F)},
\]
with
\[
\mathbf{x}_{\mathrm{leo\text{-}clk}}^{(l)}
\!\!=\!\!
\big[ \!
\Delta t_l,
\Delta\dot{t}_l,
\delta_{l,1},\dots,\delta_{l,F},
b_{l,1},\dots,b_{l,F}
\! \big]^{T} \!\! \in \! \mathbb{R}^{2+2F}.
\]

\item \textbf{GNSS clock and hardware–related states:}
GNSS satellite clock offsets and drifts, together with transmitter carrier-phase and pseudorange hardware biases.
\[
\mathbf{x}_{\mathrm{gnss\text{-}clk}}
=
\operatorname{col}\big\{
\mathbf{x}_{\mathrm{gnss\text{-}clk}}^{(1)},\;
\ldots,\;
\mathbf{x}_{\mathrm{gnss\text{-}clk}}^{(g)}
\big\}
\in\mathbb{R}^{G(2+2F)},
\]
with
\[
\mathbf{x}_{\mathrm{gnss\text{-}clk}}^{(g)} \!
= \!
\big[
\Delta t^{g},
\Delta\dot{t}^{g},
\delta^{g}_{1},\dots,\delta^{g}_{F},
b^{g}_{1},\dots,b^{g}_{F}
\big]^{T} \! \in \! \mathbb{R}^{2+2F}.
\]
\item \textbf{Ionosphere-related states:}
ionospheric delays associated with each observable LEO–GNSS link.
\[
\mathbf{x}_{\mathrm{ion}}
=
\operatorname{col}\big\{
\mathbf{x}_{\mathrm{ion}}^{(1)},
\ldots,
\mathbf{x}_{\mathrm{ion}}^{(L)}
\big\}
\in \mathbb{R}^{\sum_l G_l},
\]
where
\[
\mathbf{x}_{\mathrm{ion}}^{(l)}
=
\big[
I_l^{g_1},\,
\ldots,\,
I_l^{g_{G_l}}
\big]^{T}
\in\mathbb{R}^{G_l}.
\]
\item \textbf{Carrier-phase ambiguity states:}
carrier-phase ambiguities for all LEO–GNSS links and frequencies.
\[
\mathbf{x}_{\mathrm{amb}}
=
\operatorname{col}\big\{
\mathbf{x}_{\mathrm{amb}}^{(1)},\;
\ldots,\;
\mathbf{x}_{\mathrm{amb}}^{(L)}
\big\}
\in \mathbb{R}^{\sum_l F G_l},
\]
with
\[
\mathbf{x}_{\mathrm{amb}}^{(l)}
=
\operatorname{col}\big\{
\mathbf{x}_{\mathrm{amb}}^{(l,1)},\;
\ldots,\;
\mathbf{x}_{\mathrm{amb}}^{(l,F)}
\big\}
\in \mathbb{R}^{F G_l},
\]
\[
\mathbf{x}_{\mathrm{amb}}^{(l,f)}
=
\begin{bmatrix}
z_{l,f}^{g_1},
\cdots,
z_{l,f}^{g_{G_l}}
\end{bmatrix}^{T}
\in\mathbb{R}^{G_l}.
\]
\end{itemize}

The vector of unknown parameters is then represented as
\begin{equation}
\mathbf{x} =
\left[
\mathbf{x}_{\mathrm{leo\text{-}orb}}^\mathrm{T},
\mathbf{x}_{\mathrm{leo\text{-}clk}}^\mathrm{T},
\mathbf{x}_{\mathrm{gnss\text{-}clk}}^\mathrm{T},
\mathbf{x}_{\mathrm{ion}}^\mathrm{T},
\mathbf{x}_{\mathrm{amb}}^\mathrm{T}
\right]^\mathrm{T} \in \mathbb{R}^{n},
\end{equation}
where
\[
n
=
6L+(L+G)(2+2F)
+(1+F)\sum_{l=1}^{L}G_l.
\]
Given the definitions of $\mathbf{y}\in\mathbb{R}^{m}$ and
$\mathbf{x}\in\mathbb{R}^{n}$, the linearized network-wide observation model can be written as
\begin{equation}
\label{eq:Ey_net}
\mathrm{E}\{\mathbf{y}\} = \mathbf{A}\mathbf{x},
\end{equation}
where $\mathbf{A}\in\mathbb{R}^{m\times n}$ is the network-wide design matrix.
Following the same five-state grouping introduced for $\mathbf{x}$, the corresponding design matrix $\mathbf{A}$ is structured as
\begin{equation}
\label{eq:A_net}
\mathbf{A} =
\left[\!\! \begin{array}{c|c|c|c|c}
\mathbf{A}_{\mathrm{leo\text{-}orb}} 
& \mathbf{A}_{\mathrm{leo\text{-}clk}}
& \mathbf{A}_{\mathrm{gnss\text{-}clk}}
& \mathbf{A}_{\mathrm{ion}}
& \mathbf{A}_{\mathrm{amb}}
\end{array} \!\! \right].
\end{equation}
The submatrices appearing in \eqref{eq:A_net} are defined in Appendix~\ref{appendix:A}.

The GNSS satellite-specific parameters $\mathbf{x}_{\mathrm{gnss\text{-}clk}}$ constitute shared global states, while all remaining parameters are receiver-related and treated as local states. This local–global parameter structure naturally leads to a network-driven precise estimation formulation, in which the global states are inferred cooperatively across the LEO constellation. The resulting global states provide a consistent reference that supports precise determination of the local LEO orbit and clock states, namely $\mathbf{x}_{\mathrm{leo\text{-}orb}}$ for LEO orbit determination and $\mathbf{x}_{\mathrm{leo\text{-}clk}}$ for constellation-wide time synchronization.
In this sense, the derived GNSS satellite clock states $\mathbf{x}_{\mathrm{gnss\text{-}clk}}$ can be interpreted as GNSS satellite clock correction information generated by the LEO network, which may serve as a potential correction source for downstream GNSS positioning applications.
The estimability of the individual parameter components, and the conditions under which they can be uniquely identified, are discussed in the next section.

\begin{table*}[b]
\centering
\begin{threeparttable}
\renewcommand{\arraystretch}{2.2}
\centering
\caption{Rank Deficiencies in Spaceborne GNSS Network: Types, Sizes, and the $\mathbf{S}$-basis Constraints Chosen for Their Elimination}
\label{tab:Rank_def}
\begin{tabularx}{\textwidth}{l|l|l|l}
\hline 
\textbf{Rank-deficiency Type} & \textbf{Size} & \textbf{Chosen $\mathbf{S}$-basis Constraint}  &\textbf{Notation}\\
\hline 

Between LEO receiver and GNSS satellite clocks & 1 & Pivot receiver clock &$\Delta t_1$\\
\hline 

Between LEO receiver and GNSS satellite clock drift  & 1 & Pivot receiver clock drift  &$\Delta \dot{t}_1$\\
\hline 

Between LEO receiver and GNSS satellite hardware biases &$2F$ &Pivot receiver hardware biases &$\left\{\begin{array}{l}\delta_{1, f} \\ b_{1, f}\end{array} f \geq 1\right.$\\
\hline

Between LEO receiver clocks and receiver hardware biases &$L-1$ &Ionosphere-free receiver code biases  &$b_{l, \mathrm{IF}}, l \geq 2$\\
\hline

Between GNSS satellite clocks and hardware biases &$G$  &Ionosphere-free satellite code biases &$b_{\mathrm{IF}}^g, g \geq 1$ \\
\hline

Between LEO receiver hardware biases and ionospheric delays  &$L-1$  &Geometry-free receiver code biases  &$b_{l, \mathrm{GF}}, l \geq 2$\\
\hline

Between GNSS satellite hardware biases and ionospheric delays &$G$  & Geometry-free satellite code biases &$b_{\mathrm{GF}}^g, g \geq 1$\\
\hline

Between LEO receiver hardware biases and ambiguities 
& $F(L-1)$ 
& Carrier-phase ambiguities for pivot satellites 
& $z_{l,f}^{g_p(l)},\; l \geq 2,\; f \geq 1$\\
\hline

Between GNSS satellite hardware biases and ambiguities 
& $FG$ 
& Carrier-phase ambiguities for pivot receivers 
& $z_{l_p(g), f}^{g},\; g \geq 1,\; f \geq 1$\\
\hline

\end{tabularx}
\begin{tablenotes}
\item[*] $b_{l, \mathrm{IF}}=\frac{\mu_2}{\mu_2-\mu_1} b_{l, 1}-\frac{\mu_1}{\mu_2-\mu_1} b_{l, 2}, \quad b_{\mathrm{IF}}^g =\frac{\mu_2}{\mu_2-\mu_1} b_{1}^g-\frac{\mu_1}{\mu_2-\mu_1} b_{2}^g, \quad b_{l, \mathrm{GF}}=\frac{1}{\mu_2-\mu_1}\left[b_{l, 2}-b_{l, 1}\right], \quad b_{\mathrm{GF}}^g =\frac{1}{\mu_2-\mu_1}\left[b_2^g-b_1^g\right]$
\end{tablenotes}
\end{threeparttable}
\end{table*}

\begin{table*}[ht]
\setlength{\tabcolsep}{1.0mm}
\renewcommand{\arraystretch}{2.2}
\centering
\caption{Estimable network parameters and their interpretation using the chosen $\mathbf{S}$-basis}
\label{tab:EstPar}
\resizebox{\textwidth}{!}{
\begin{tabular}{l|l|l}
\hline 
\textbf{Estimable parameter} & \textbf{Notation and interpretation} & \textbf{Condition}\\
\hline 

LEO receiver position &$ \Delta \tilde{\mathbf{p}}_l = \Delta \mathbf{p}_l$  &$l \geq 1$\\
\hline 

LEO receiver velocity &$ \Delta \tilde{\mathbf{v}}_l = \Delta \mathbf{v}_l$ &$l \geq 1$\\
\hline

LEO receiver clock &$\Delta \tilde{t}_l = \left[\Delta t_l + b_{l, \mathrm{IF}}\right]-\left[\Delta t_1+b_{1, \mathrm{IF}}\right] $ &$l \geq 2$\\
\hline 

LEO receiver clock drift &   $\Delta \tilde{\dot{t}}_l = \Delta \dot{t}_l - \Delta \dot{t}_1$    &$l \geq 2$\\
\hline

LEO receiver phase bias  &$\tilde{\delta}_{l, f}=\left[\delta_{l, f}-\frac{1}{\lambda_f} b_{l, \mathrm{IF}}\right]-\left[\delta_{1, f}-\frac{1}{\lambda_f} b_{1, \mathrm{IF}}\right]+\frac{1}{\lambda_f} \mu_f\left[b_{l, \mathrm{GF}}-b_{1, \mathrm{GF}}\right]+\left[z_{l, f}^{g_p(l)}-z_{l_p(g), f}^{g_p(l)}\right]$  &$l \geq 2, f \geq 1$\\
\hline

LEO receiver code bias  &$\tilde{b}_{l, f} = \left[b_{l, f}-b_{l, \mathrm{IF}}\right]-\left[b_{1, f}-b_{1, \mathrm{IF}}\right]-\mu_f\left[b_{l, \mathrm{GF}}-b_{1, \mathrm{GF}}\right]$ &$l \geq 2, f \geq 3$\\
\hline

GNSS satellite clock  &$\Delta \tilde{t}^g=\left[\Delta t^g+b_{\mathrm{IF}}^g\right]-\left[\Delta t_1+b_{1, \mathrm{IF}}\right]$  &$g \geq 1$ \\
\hline

GNSS satellite clock drift & $\Delta \tilde{\dot{t}}^g = \Delta \dot{t}^g - \Delta \dot{t}_1$       &$g \geq 1$  \\
\hline

GNSS satellite phase bias & $\tilde{\delta}_j^g=\left[\delta_f^g-\frac{1}{\lambda_f} b_{\mathrm{IF}}^g\right]-\left[\delta_{1, f}-\frac{1}{\lambda_f} b_{1, \mathrm{IF}}\right] + \frac{1}{\lambda_f} \mu_f\left[b_{\mathrm{GF}}^g-b_{1, \mathrm{GF}}\right] - z_{l_p(g),f}^g$   & $g \geq 1, f \geq 1$\\
\hline

GNSS satellite code bias &$\tilde{b}_f^g = \left[b_f^g-b_{\mathrm{IF}}^g\right]-\left[b_{1, f}-b_{1, \mathrm{IF}}\right]-\mu_f\left[b_{\mathrm{GF}}^g-b_{1, \mathrm{GF}}\right]$  &$g \geq 1, f \geq 3$\\
\hline

Ionospheric delay &$\tilde{l}_l^g =l_l^g + b_{l, \mathrm{GF}}-b_{\mathrm{GF}}^g$ &$l \geq 1, g \geq 1$\\
\hline

Carrier-phase ambiguity &$\tilde{z}_{l, f}^g=\left[z_{l, f}^g-z_{l, f}^{g_p(l)}\right]-\left[z_{l_p(g),f}^g-z_{l_p(g), f}^{g_p(l)}\right] $ & $l \neq  l_p(g), g \neq g_p(l), f \geq 1$\\
\hline
\end{tabular}
}
\end{table*}

\section{Estimability and Identifiability Analysis of Model Parameters}
\label{sec:estimability}

The observation equations introduced in Section~\ref{sec:observation} inherently lead to a rank-deficient estimation problem: the design matrix $\mathbf{A}$ is not full rank, meaning that only certain linear combinations of the parameters are identifiable, while their absolute values are not. For a networked architecture, this issue is particularly critical, as all nodes must adopt a common set of physically meaningful estimable parameters; otherwise, different nodes would converge to mutually inconsistent solutions. To address this challenge, we adopt the $\mathcal{S}$-system theory to rigorously characterize the identifiable parameter combinations of the proposed spaceborne GNSS network~\cite{baarda1973s,teunissen1985zero,odijk2016estimability}.

To identify the rank deficiencies of the design matrix in \eref{eq:Ey_net}, we examine which changes in the unknown parameters leave all measurements unchanged. A straightforward example is a common shift applied simultaneously to all receiver clocks and all GNSS satellite clocks, which does not affect either the carrier-phase or pseudorange observations. It indicates that the corresponding clock parameters are not independently observable. Such a dependency introduces a rank deficiency, which can be resolved by designating one receiver clock as the temporal reference, thereby removing it from the set of unknowns. This constraint forms part of the adopted $\mathbf{S}$-basis.

Applying the same reasoning to the full model reveals additional parameter dependencies. These dependency categories, together with the specific constraints used to eliminate them, are summarized in \tref{tab:Rank_def}. As defined previously, $\mathbf{y}\in\mathbb{R}^{m}$, $\mathbf{x}\in\mathbb{R}^{n}$, and the corresponding design matrix satisfies $\mathbf{A}\in\mathbb{R}^{m\times n}$, with $\operatorname{rank}(\mathbf{A})=r\leq n$. The total dimension of the associated null space, i.e., the cumulative size of all rank deficiencies, is
\begin{equation}
n-r = 2 + 2F + (2+F)(L-1+G).
\end{equation}

The range and null spaces of the matrix $\mathbf{A}$ are denoted by $\mathcal{R}(\mathbf{A})$ and $\mathcal{N}(\mathbf{A})$, with dimensions $\operatorname{dim}\mathcal{R}(\mathbf{A}) = r$ and $\operatorname{dim}\mathcal{N}(\mathbf{A}) = n - r$, respectively. Let $\mathbf{V} \in \mathbb{R}^{n \times (n-r)}$ be a matrix whose columns form a basis of the null space $\mathcal{N}(\mathbf{A})$, so that $\mathbf{A}\mathbf{V} = \mathbf{0}$. Although the null space $\mathcal{N}(\mathbf{A})$ is unique, the particular choice of its basis matrix $\mathbf{V}$ is not. 
Let $\mathbf{S} \in \mathbb{R}^{n \times r}$ denote a basis matrix spanning a subspace complementary to $\mathcal{R}(\mathbf{V})$, such that
\[
\mathbb{R}^n = \mathcal{R}(\mathbf{S}) \oplus \mathcal{R}(\mathbf{V}).
\]

Because the matrix $\mathbf{A}$ is rank-deficient, i.e., $r < n$, the observations do not contain sufficient information to determine all components of the parameter vector. In this case, $\mathbf{x}$ can be decomposed into two parts: an estimable component $\mathbf{x}_{_{\mathbf{S}}} \in \mathcal{R}(\mathbf{S})$ and a non-estimable component $\mathbf{x}_{_{\mathbf{V}}} \in \mathcal{R}(\mathbf{V})$. This decomposition can be written as
\begin{equation}
\mathbf{x}
= \underbrace{\mathbf{S}\alpha}_{\mathbf{x}_{_{\mathbf{S}}}}
+ \underbrace{\mathbf{V}\beta}_{\mathbf{x}_{_{\mathbf{V}}}}
= 
\underbrace{\begin{bmatrix}\mathbf{S} & \mathbf{V}\end{bmatrix}}_{n \times n}
\begin{bmatrix}
\alpha \\
\beta
\end{bmatrix},
\end{equation}
where $\alpha \in \mathbb{R}^{r}$ contains the estimable parameter functions associated with $\mathbf{S}$, and $\beta \in \mathbb{R}^{\,n-r}$ contains the inestimable parameter functions associated with $\mathbf{V}$.

Since the square matrix $\begin{bmatrix}\mathbf{S} & \mathbf{V}\end{bmatrix} \in \mathbb{R}^{n \times n}$ is invertible, the coefficients $\alpha$ and $\beta$ can be obtained from
\begin{equation}
\begin{bmatrix}
\alpha \\[1mm]
\beta
\end{bmatrix}
=
\begin{bmatrix}
\mathbf{S} & \mathbf{V}
\end{bmatrix}^{-1}\mathbf{x}
=
\begin{bmatrix}
\left[\left(\mathbf{V}^{\perp}\right)^{ T}\mathbf{S}\right]^{-1}\left(\mathbf{V}^{\perp}\right)^{ T} \\[2mm]
\left[\left(\mathbf{S}^{\perp}\right)^{ T}\mathbf{V}\right]^{-1}\left(\mathbf{S}^{\perp}\right)^{ T}
\end{bmatrix}
\mathbf{x}.
\end{equation}
Here, $\mathbf{V}^{\perp} \in \mathbb{R}^{n \times r}$ is a basis matrix whose columns span a subspace orthogonal to $\mathcal{R}(\mathbf{V})$, satisfying $(\mathbf{V}^{\perp})^{T}\mathbf{V} = \mathbf{0}$. 
Similarly, $\mathbf{S}^{\perp} \in \mathbb{R}^{n \times(n-r)}$ is defined such that $(\mathbf{S}^{\perp})^{T}\mathbf{S} = \mathbf{0}$.

Based on the chosen basis matrices $\mathbf{S}$ and $\mathbf{V}$, the observation model becomes a full-rank system in the estimable parameters $\alpha$:
\begin{equation}
\mathrm{E}\{\mathbf{y}\}=\mathbf{A}(\mathbf{S}\alpha + \mathbf{V}\beta) = \underbrace{(\mathbf{AS})}_{\tilde{\mathbf{A}}} \alpha.
\end{equation}
This identifies the estimable component of $\mathbf{x}$ as
\begin{equation}
\mathbf{x}_{\mathbf{S}} = \mathbf{S}\alpha
= \mathbf{x} - \mathbf{V}\beta
= \mathcal{S}\mathbf{x},
\end{equation}
where $\mathcal{S} \in \mathbb{R}^{n \times n}$ denotes the 
$\mathcal{S}$-transformation matrix. It admits the equivalent
representations
\begin{equation}
\begin{aligned}
\mathcal{S} &=
\mathbf{S} \left[(\mathbf{V}^{\perp})^{T}\mathbf{S}\right]^{-1}
(\mathbf{V}^{\perp})^{T} \\[1mm]
&= \mathbf{I}_n
- \mathbf{V} \left[(\mathbf{S}^{\perp})^{T}\mathbf{V}\right]^{-1}
(\mathbf{S}^{\perp})^{T}.
\end{aligned}
\end{equation}
The matrix $\mathcal{S}$ therefore defines the linear combinations of the initial parameters that can be estimated based on the chosen $\mathbf{S}$-basis.

Different choices of the $\mathbf{S}$-basis may be adopted, with each choice defining a particular set of estimable global products and LEO-specific states. \tref{tab:Rank_def} presents the selection of $\mathbf{S}$-basis constraints used in this work to eliminate the rank deficiencies. The corresponding matrices $\mathbf{S}$, $\mathbf{V}$, $(\mathbf{S}^{\perp})^{T}$, and $\mathcal{S}$ are provided in Appendix~\ref{appendix:B}, where they are organized into five column blocks consistent with the five parameter groups in the network model. Then, we can obtain a full-rank, undifferenced spaceborn GNSS network model. The resulting parameters are no longer the original absolute states but their estimable linear combinations, denoted by the symbol $\tilde{(\cdot)}$. Table~\ref{tab:EstPar} summarizes these estimable network parameters and clarifies their interpretation under the adopted $\mathbf{S}$-basis following the corresponding $\mathcal{S}$-transformation. The carrier-phase ambiguities are transformed into a double-differenced form, in which the initial phase terms are eliminated and the resulting ambiguities are integer-valued, thereby enabling integer-constrained high-precision positioning.

Please note that, for simplicity, the epoch index has been omitted in the above formulations. Following standard multi-epoch undifferenced GNSS modeling, the observation equations can be stacked over multiple epochs, where time variation enters through changing geometry and satellite visibility. 
Importantly, the $\mathbf{S}$-basis identified from a single-epoch formulation applies equally to the stacked multi-epoch model and fully characterizes the set of estimable parameter combinations. To avoid unnecessary complexity, an explicit multi-epoch formulation is not provided here. The objective of this section is to present the fundamental methodology for establishing a well-defined estimation problem under the proposed spaceborne GNSS network model, which underpins the decentralized estimation framework developed later.

\section{Decentralized Estimation for LEO Onboard GNSS Processing}
\label{sec:formulation_algorithm}
In this section, we introduce a networked onboard processing framework for LEO constellations. Building on the analysis in the previous sections, we derive a well-defined decentralized problem formulation in which all nodes estimate a common set of identifiable parameters, enabling consistent network-wide estimation for onboard processing. We then employ a momentum-accelerated gradient tracking (GT) algorithm to solve the resulting problem. The proposed framework is tailored to the orbital dynamics and inter-satellite communication constraints of LEO constellations.

\subsection{Well-defined Decentralized Formulation}
\label{sec:formulation}

Upon the full-rank design matrix from Section~\ref{sec:estimability}, the identifiable parameter vector admits a centralized weighted least-squares formulation, which serves as a natural reference solution:
\begin{equation} 
\min _{\tilde{\mathbf{x}}}\|\mathbf{y}-\tilde{\mathbf{A}} \tilde{\mathbf{x}}\|^2_{\mathbf{Q}^{-1}},
\end{equation}
where $\mathbf{Q}$ denotes the observation-noise covariance matrix.

However, such a centralized architecture is challenging for large-scale LEO constellations.
Currently, no dedicated spaceborne computing center exists, and deploying and operating such an infrastructure would entail substantial cost and system complexity. Even if a fusion node were available, transmitting raw observations to a central processor and forming global normal equations does not scale with the constellation size or observation duration, and would incur prohibitive communication overhead, increased latency, and limited robustness. These limitations motivate a decentralized formulation, in which each LEO node performs local computation while exchanging only compact information with its neighboring satellites.

Concretely, we introduce an equivalent yet scalable decentralized formulation:
\begin{equation}
\begin{aligned}
& \underset{\{\mathbf{x}_l\}^{L}_{l=1},\ \mathbf{z}}{\text{minimize}} \quad f(\{\mathbf{x}_l\}^{L}_{l=1},\mathbf{z})=\sum_{l=1}^{L}f_{l}(\mathbf{x}_{l},\mathbf{z}),\\
 &\mathrm{where} \quad f_{l}(\mathbf{x}_{l},\mathbf{z}) = \frac{1}{2}\big\|\,\mathbf{A}_{l}\mathbf{x}_{l}+\mathbf{B}_{l}\mathbf{z}-\mathbf{y}_l\,\big\|_{\mathbf{Q}_{l}^{-1}}^2,
\end{aligned}
\label{eq:obs_dec}
\end{equation}
where $\mathbf{A}_l$ and $\mathbf{B}_l$ are design matrices relating observations at node $l$ to its node-specific parameter vector $\mathbf{x}_l$
and the shared global parameter vector $\mathbf{z}$, respectively,
and $\mathbf{Q}_l$ denotes the local observation-noise covariance
matrix. Specifically, $\mathbf{A}_l$ and $\mathbf{B}_l$ are obtained by partitioning the full-rank design matrix $\tilde{\mathbf{A}}$ into blocks corresponding to node-specific and globally shared parameters, respectively. This partition follows from the estimability analysis in Section~\ref{sec:estimability},
with a common $\mathbf{S}$-basis defining a consistent identifiable parameterization of $f_l(\mathbf{x}_l,\mathbf{z})$ in a physically meaningful manner.
The resulting formulation retains the physical structure of
the observation model: node-specific parameters are estimated
locally, whereas shared estimable quantities require agreement
across the network.
\begin{remark}[Estimability-guided decentralized formulation]
Generic decentralized optimization frameworks typically take the local objective functions or data matrices as given and impose suitable regularity conditions. In \eqref{eq:obs_dec}, the local objectives $f_l(\mathbf{x}_l,\mathbf{z})$ are constructed from the estimability analysis of the physical observation model. The partition into node-specific variables $\mathbf{x}_l$ and shared variables $\mathbf{z}$ follows from the identifiable parameterization and determines the structure of $\mathbf{A}_l$ and $\mathbf{B}_l$. Specifically, the selected $\mathbf{S}$-basis defines the estimable parameter combinations represented by $\mathbf{x}_l$ and $\mathbf{z}$. Different admissible $\mathbf{S}$-bases preserve the observable quantities and total estimable dimension, but may change the local--shared partition and the dimension of $\mathbf{z}$. For example, with other rank deficiencies resolved, the clock datum may be defined by a pivot LEO receiver or a reference GNSS satellite. The clock parameters are then expressed relative to the selected reference, while the receiver--satellite clock differences entering the observations remain unchanged. A common $\mathbf{S}$-basis ensures that all nodes impose consensus on the same parameter combinations with respect to the same datum. However, this consistency does not imply optimal numerical conditioning. Different admissible bases may change the condition number of the reduced objective, thereby affecting the convergence rate and stepsize bounds of iterative solvers. The $\mathbf{S}$-basis adopted here is chosen to establish a well-defined, common full-rank estimation problem whose estimable parameters have a physical interpretation and provide the required correction information.
\end{remark}

\subsection{Momentum-Accelerated Gradient Tracking over Time-Varying Graphs}

\begin{figure*}[t]
  \centering
  \includegraphics[width=\textwidth]{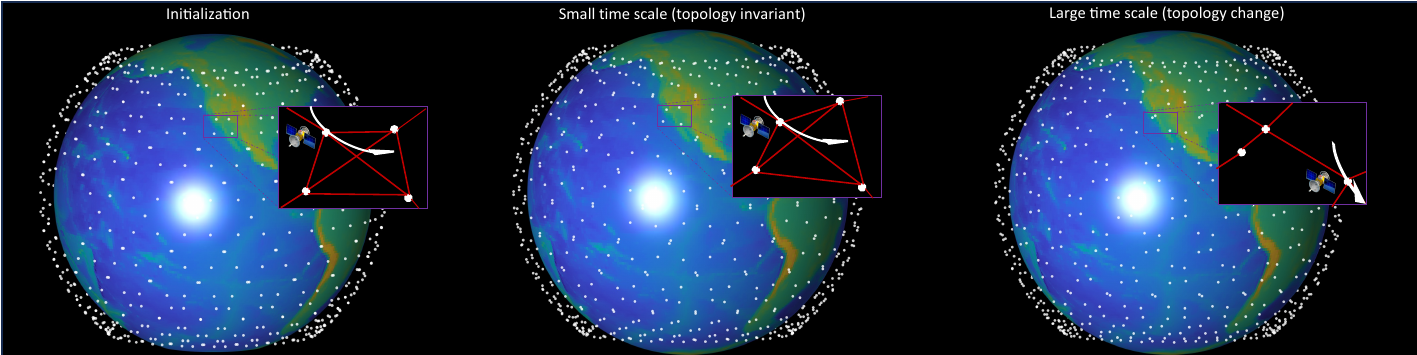}
  \caption{LEO constellation communication topology across two time scales. Over short horizons, the topology is effectively invariant as links persist despite satellite motion; over longer horizons, the topology changes as satellites move out of range of previous neighbors and establish new links. This motivates modeling the optimization over a piecewise-invariant graph sequence $\{\mathcal{G}^{(t)}\}^{T}_{t=1}$.}
  \label{fig:jsac}
\end{figure*}

We solve \eqref{eq:obs_dec} using a Heavy-ball GT scheme based on the GT framework~\cite{nedic2017achieving} and momentum-accelerated GT methods~\cite{xin2019distributed,huang2026accelerated}. The scheme is adapted to the estimability-guided objective $f_l(\mathbf{x}_l,\mathbf{z})$ and the communication dynamics of LEO constellations. Momentum is used to accelerate the iterations, while multi-round consensus reduces inter-satellite disagreement through repeated neighborhood mixing. As illustrated in Fig.~\ref{fig:jsac}, we characterize the inter-satellite communication topology of LEO constellations as exhibiting distinct behaviors across two time scales: on a small time scale, the network topology remains effectively invariant, as communication links persist despite the continuous motion of the satellites; on a large time scale, the topology evolves as satellites gradually drift out of range of previous neighbors and establish new connections. Consequently, since the constellation operates iteratively through these regimes, we model the optimization scheme over a piecewise-invariant graph.

Let us assume the optimization of \eqref{eq:obs_dec} spans a sequence of graphs $\{\mathcal{G}^{(t)}=(\mathcal{V},\mathcal{E}^{(t)},\mathbf{W}^{(t)})\}^{T}_{t=1}$, where each $\mathcal{G}^{(t)}$ dominates a specific time period. Here, $\mathcal{G}^{(t)}$ represents a topology-invariant graph snapshot: although the satellites move continuously, the active inter-satellite links remain unchanged within the corresponding interval. The discrete-time optimization samples the connectivity evolution at discrete epochs and uses the fixed mixing matrix $\mathbf{W}^{(t)}$ throughout each interval. The graph is updated only when the connectivity pattern changes. Note that $T$ denotes the number of topology-invariant intervals over the considered horizon, whereas the number of optimization updates that can be performed within each interval depends on practical factors such as onboard computing capability, communication latency, and available bandwidth. 
\begin{remark}[Physical interpretation of the two-timescale abstraction]
Continuous satellite motion changes the inter-satellite link geometry but does not necessarily change the communication topology. Inter-satellite links can be established and maintained by pointing, acquisition, and tracking (PAT) systems: ephemerides and attitude information support initial pointing, while tracking compensates for the changing line-of-sight direction after acquisition. Consequently, the communicating neighbors can remain unchanged over an interval even as the satellites move. The two-timescale abstraction applies to operating regimes in which each optimization iteration, including its local computations and required inter-satellite exchanges, can be completed within such a topology-invariant interval. On the short timescale, optimization iterations and their consensus rounds use the unchanged topology $\mathcal{G}^{(t)}$ and its mixing matrix $\mathbf{W}^{(t)}$. On the long timescale, changes in the neighbor sets, as existing links are lost or new links are established, define transitions from $\mathcal{G}^{(t)}$ to $\mathcal{G}^{(t+1)}$. Thus, the abstraction separates the timescale of iterative computation and communication from that of topology changes; it does not require fixed satellite positions or link geometry. At each estimation epoch, the observations and objective function are held fixed during the iterative solution, so convergence concerns a fixed estimation problem over the considered piecewise-invariant graph sequence.
\end{remark}

Therefore, let $L$ LEO satellites form the node set $\mathcal{V}=\{1,\ldots,L\}$, while the edge set $\mathcal{E}^{(t)} \subseteq \mathcal{V} \times \mathcal{V}$ represents time-varying inter-satellite communication links. The mixing matrix
$\mathbf{W}^{(t)}=[w^{(t)}_{lq}]\in\mathbb{R}^{L\times L}$ is compatible with the topology, i.e.,
$w^{(t)}_{lq}>0$ only if $(l,q)\in\mathcal{E}^{(t)}$ (or $l=q$), and we assume $\mathbf{W}^{(t)}$ is doubly stochastic:
$\mathbf{W}^{(t)}\mathbf{1}=\mathbf{1}$ and $\mathbf{1}^{\top}\mathbf{W}^{(t)}=\mathbf{1}^{\top}$. We adopt Metropolis weights for $\mathbf{W}^{(t)}$, which can be computed in a decentralized manner by exchanging node-degree information.
Specifically, letting $\mathcal{N}^{(t)}_l=\{q\in\mathcal{V}:(l,q)\in\mathcal{E}^{(t)}\}$ denote the neighbor set of node $l$
and $d^{(t)}_l = |\mathcal{N}^{(t)}_l|$ its degree, the weights are 
\begin{equation}
w_{lq}^{(t)} = 
\begin{cases} 
\frac{1}{\max\left(d_l^{(t)}, d_q^{(t)}\right) + 1} & \text{if } q \in \mathcal{N}_l^{(t)}, \\
1 - \sum_{p \in \mathcal{N}_l^{(t)}} w_{lp}^{(t)} & \text{if } l = q, \\
0 & \text{otherwise}.
\end{cases}
\end{equation}

Then, for iteration $k$ within the time period dominated by $\mathcal{G}^{(t)}$, the algorithm at node $l$ proceeds as follows
\begin{subequations}\label{eq:update_rules}
\begin{align}
\mathbf{v}_l^{k} &= \mathbf{z}_l^{k} + \theta \left( \mathbf{z}_l^{k} - \mathbf{z}_l^{k-1} \right), \label{eq:update_rules:a}\\
\mathbf{z}_l^{k+1} &= \mathcal{C}\!\left(\mathbf{v}_l^{k} - \gamma \mathbf{g}_l^{k},\,\mathcal{K}\right), \label{eq:update_rules:b}\\
\mathbf{x}_l^{k+1} &=
(\mathbf{A}_l^\top \mathbf{Q}_{l}^{-1}\mathbf{A}_l)^{-1}\mathbf{A}_l^\top \mathbf{Q}_{l}^{-1}
\bigl(\mathbf{y}_l-\mathbf{B}_l\mathbf{z}_l^{k+1}\bigr), \label{eq:update_rules:c}\\
\mathbf{g}_{l}^{k+1} &=
\sum_{q\in\mathcal{N}^{(t)}_l} w^{(t)}_{ql}\,\mathbf{g}_q^k
+ \nabla_{\mathbf{z}} f_{l}(\mathbf{x}_{l}^{k+1},\mathbf{z}_{l}^{k+1})
-\nabla_{\mathbf{z}} f_{l}(\mathbf{x}_{l}^{k},\mathbf{z}_{l}^{k}), \label{eq:update_rules:d}
\end{align}
\end{subequations}
where $\nabla_{\mathbf{z}} f_{l}(\mathbf{x}_{l},\mathbf{z}_{l})$ denotes the gradient of $f_l(\mathbf{x}_{l},\mathbf{z}_{l})$ with respect to $\mathbf{z}$.

Under this formulation, \eqref{eq:update_rules:a} implements the Heavy-ball momentum update $\mathbf{v}_l^{k}$ with coefficient $\theta \in [0,1)$; \eqref{eq:update_rules:b} performs a $\gamma$ descent step combined with a $\mathcal{K}$-round consensus operator $\mathcal{C}(\cdot)$ to obtain $\mathbf{z}_l^{k+1}$; \eqref{eq:update_rules:c} computes the closed-form update of the local auxiliary variable $\mathbf{x}_l$ given $\mathbf{z}_l^{k+1}$ (as derived in \cite{zheng2025decentralized}); and \eqref{eq:update_rules:d} updates the gradient tracker $\mathbf{g}_l^{k+1}$. The complete iterative procedure is summarized in Algorithm~\ref{alg:gt-atc}. For simplicity, we assume that each snapshot $\mathcal{G}^{(t)}$ remains dominant for the same period length.

The consensus operator $\mathcal{C}(\cdot)$ performs $\mathcal{K}$ rounds of neighborhood mixing using the mixing matrix $\mathbf{W}^{(t)}$. 
The use of multiple consensus rounds is motivated by LEO constellations, where the inter-satellite communication graph is typically sparse (e.g., each satellite may communicate with only a few neighbors), which may limit information mixing in a single round. 
By incorporating multi-round mixing, $\mathcal{C}(\cdot)$ strengthens information diffusion across the network and can stabilize the updates, allowing a larger momentum parameter $\theta$ and stepsize $\gamma$ and thereby improving convergence in practice.

\begin{algorithm}[t]
\caption{Momentum-Accelerated GT for Decentralized Onboard GNSS Processing in LEO Constellations}
\label{alg:gt-atc}
\begin{algorithmic}[1]
\Require Graph patterns $\{\mathcal{G}^{(t)}\}^{T}_{t=1}$; stepsize $\gamma>0$; momentum $\theta$; maximum iterations $K$, consensus rounds $\mathcal{K}$
\Require Local data $\{\mathbf{A}_l,\mathbf{B}_l,\mathbf{Q}_{l},\mathbf{y}_l\}_{l=1}^L$ and $f_l$ as in \eqref{eq:obs_dec}
\Ensure Estimates $\{\mathbf{x}_l^{K},\mathbf{z}_l^{K}\}_{l=1}^L$
\State \textbf{Initialization:} For each node $l$, initialize $\mathbf{z}_l^0=\mathbf{z}_l^{-1}$, calculate $\mathbf{x}_l^0 \gets (\mathbf{A}_l^\top \mathbf{Q}_{l}^{-1}\mathbf{A}_l)^{-1}\mathbf{A}_l^\top \mathbf{Q}_{l}^{-1}(\mathbf{y}_l-\mathbf{B}_l\mathbf{z}_l^0)$, and set $\mathbf{g}_l^0 \gets \nabla_{\mathbf{z}} f_l(\mathbf{x}_l^0,\mathbf{z}_l^0)$ 
\For{$k=0$ \textbf{to} $K-1$}
  \State $t \gets \lfloor k / T\rfloor+1$ ;
  \State $\mathbf{W}^{(t)}\gets\mathcal{G}^{(t)}$ ; 
  \ForAll{$l \in \{1,\ldots,L\}$} 
  \State $\mathbf{v}_l^{k} \gets \mathbf{z}_l^{k} + \theta \left( \mathbf{z}_l^{k} - \mathbf{z}_l^{k-1} \right)$ ;
  \State $\boldsymbol{\psi}_l^k\gets\mathbf{v}_l^{k} - \gamma \mathbf{g}_l^{k}$ ;
  \State $\boldsymbol{\varphi}_l\gets\boldsymbol{\psi}_l^k$ ;
   \For{$\kappa=0$ \textbf{to} $\mathcal{K}-1$}
   \State $\boldsymbol{\varphi}_l \gets \sum_{q\in\mathcal{N}^{(t)}_l} w^{(t)}_{ql}\boldsymbol{\varphi}_q$ ;
   \EndFor
    \State $\mathbf{z}_l^{k+1} \gets \boldsymbol{\varphi}_l$ ;
    \State $\mathbf{x}_l^{k+1} \gets (\mathbf{A}_l^\top \mathbf{Q}_{l}^{-1}\mathbf{A}_l)^{-1}\mathbf{A}_l^\top \mathbf{Q}_{l}^{-1}(\mathbf{y}_l-\mathbf{B}_l\mathbf{z}_l^{k+1})$ ;
    \State $\mathbf{g}_l^{k+1} \gets \sum_{q\in\mathcal{N}^{(t)}_l} w^{(t)}_{ql}\,\mathbf{g}_q^k + \nabla_{\mathbf{z}} f_{l}(\mathbf{x}_{l}^{k+1},\mathbf{z}_{l}^{k+1})-\nabla_{\mathbf{z}} f_{l}(\mathbf{x}_{l}^{k},\mathbf{z}_{l}^{k})$ ;
  \EndFor
\EndFor
\end{algorithmic}
\end{algorithm}

\begin{remark}[Connectivity and convergence conditions]
We assume that each nominal inter-satellite graph $\mathcal{G}^{(t)}$ is undirected and connected. The corresponding Metropolis matrix $\mathbf{W}^{(t)}$ is symmetric and doubly stochastic, and its non-dominant eigenvalues have magnitude strictly less than one. Thus, mixing preserves the network average and contracts disagreement. Within a topology-invariant interval, $\mathcal{K}$ consensus rounds reduce disagreement by a factor bounded by $(\rho_t)^{\mathcal{K}}$, where $\rho_t=\|\mathbf{W}^{(t)}-\mathbf{J}\|_2<1$ and $\mathbf{J}=\mathbf{1}\mathbf{1}^{\top}/L$. Across connected nominal graphs, the contraction of a sequence of mixing steps is bounded by the product of their individual contraction factors. Thus, the mixing rate depends on the graphs encountered and the number of exchanges, rather than on topology switching frequency alone. When the nominal topology changes, Metropolis weights are updated using the current neighbor degrees, without requiring global topology information. The estimation objective remains fixed throughout the iterative solution at each epoch. Convergence also requires smoothness and convexity of the reduced objective and suitable stepsize and momentum choices \cite{nedic2017achieving,xin2019distributed,huang2026accelerated}. If link failures temporarily disconnect the active graphs, individual mixing steps need not contract disagreement. Bounded-window joint connectivity, together with suitable mixing-weight conditions, ensures contraction over finite windows \cite{nedic2017achieving}. In \cite{zheng2025decentralized}, mixing weights are designed to reduce the block contraction factor, and convergence of the gradient-tracking scheme studied there is established under block contraction and a sufficiently small stepsize, allowing individual graphs to be disconnected.
\end{remark}

\begin{remark}[Reducing communication overhead]
Let $d_z=\mathrm{dim}(\mathbf{z})$ and let $n_l^{(t)}$ denote the number of neighbors of satellite $l$ in $\mathcal{G}^{(t)}$. Each iteration involves $\mathcal{K}$ exchanges of the consensus variable and one gradient-tracker exchange. With separate full-vector transmission to each neighbor, satellite $l$ sends $(\mathcal{K}+1)n_l^{(t)}d_z$ scalars per iteration, excluding protocol overhead. The network-wide cost scales as $\mathcal{O}((\mathcal{K}+1)|\mathcal{E}^{(t)}|d_z)$. Additional consensus rounds improve mixing but increase the per-iteration communication payload; the total communication required to reach a prescribed accuracy depends on the number of outer iterations as well. Metropolis weights are computed locally from neighbor degrees and adapt to topology changes without requiring global topology information. However, mixing over sparse graphs may be slow and require many exchanges. When coarse constellation-level ephemerides are available onboard for routing and link planning, they can also be used to predict the nominal graph sequence under the assumed link-availability conditions. This topology information enables the design of mixing coefficients beyond local rules such as Metropolis weights. The repeated exchanges within each topology-invariant interval can be viewed as a finite-impulse-response (FIR) graph filter: repeated Metropolis mixing applies $(\mathbf{W}^{(t)})^{\mathcal{K}}$, whereas designed coefficients can provide a different attenuation of disagreement while preserving the consensus component. Finite-time consensus~\cite{nguyen}, edge-variant filtering~\cite{coutino2019advances}, and cascaded graph filters~\cite{9496112} may therefore attain the required mixing accuracy with fewer communication rounds~\cite{9496112,vlaski2023networked}. These designs require additional coefficient computation, with method-dependent storage and arithmetic costs. Coefficients can be computed offline for predicted nominal topologies or updated online, and then applied through local weighted exchanges. Unanticipated topology changes require checking their applicability or updating the design. Low-bit quantization with error feedback may further reduce the transmitted bits~\cite{roula,zheng2025quantitative}. These filter and quantization extensions are not evaluated in this work. Activating fewer links also reduces communication per round, but may weaken consensus contraction and require more consensus rounds or outer iterations, so total communication need not decrease. Finally, additional consensus rounds involve no new gradient evaluations, but their effect on wall-clock time depends on communication latency and the number of outer iterations saved.
\end{remark}

\section{Numerical Results}
\label{sec:numerical_results}

Numerical simulations are conducted to evaluate the proposed network-based LEO onboard GNSS processing framework. A Walker--Delta LEO constellation consisting of 500 satellites is considered, where each satellite is equipped with a dual-frequency GNSS receiver. The satellites are deployed in 20 orbital planes with 25 satellites per plane, at an altitude of 550~km and an inclination of $53^\circ$, and their trajectories are propagated accordingly. Each LEO satellite is assumed to communicate with its 4 nearest neighbors, thereby forming a sparse, time-varying communication graph for information exchange. GNSS satellite positions are generated by propagating the corresponding orbital parameters. Dual-frequency observations at the GPS L1/L2 frequencies are simulated, including undifferenced carrier-phase, code pseudorange, and Doppler measurements. An elevation mask of $0^\circ$ is applied, retaining only LOS-visible GNSS satellites above the local horizon of each LEO receiver at each epoch. Higher masks are commonly used for ground receivers to limit atmospheric and multipath errors at low elevations. These considerations are less applicable to the spaceborne scenario considered here. We therefore retain low-elevation observations under the assumed elevation-independent noise model. A higher elevation mask would reduce the number of available observations and change the observation geometry. Measurement noise is modeled as zero-mean Gaussian, with standard deviations of 1~mm for carrier phase, 10~cm for code, and 0.5~Hz for Doppler. Receiver clock offsets are initialized with a standard deviation of 100~ns, while the GNSS satellite clock offsets are assigned an initial uncertainty of 10~ns. The resulting simulated observation set is used for the subsequent estimation analysis.

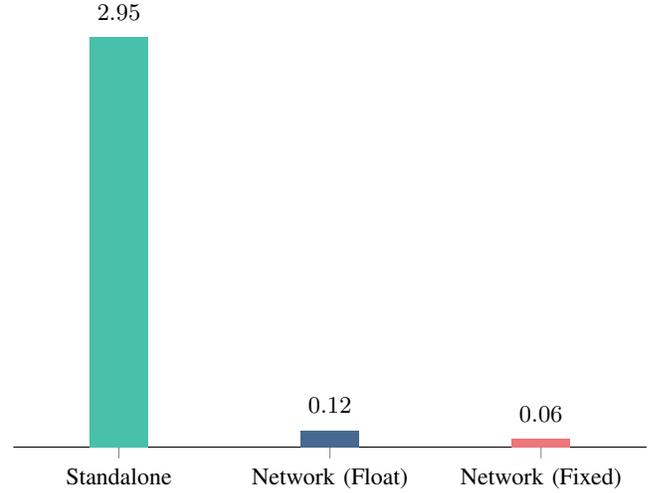
\begin{figure}[t]
\centering
\begin{tikzpicture}
\begin{axis}[
    ybar,
    bar width=22pt,
    width=10cm,
    height=7.5cm,
    ymin=0,
    ymax=3.2,
    symbolic x coords={Standalone,Network (Float),Network (Fixed)},
    xtick={Standalone,Network (Float),Network (Fixed)},
    xticklabel style={font=\small},
    ytick=\empty,
    ylabel={},
    y axis line style={draw=none},
    nodes near coords,
    nodes near coords align={vertical},
        nodes near coords style={
        /pgf/number format/fixed,
        /pgf/number format/precision=2
    },
    every node near coord/.append style={font=\small, yshift=3pt},
    axis lines=left,
    axis line style={-},
    tick align=outside,
    enlarge x limits=0.25,
    bar shift=0pt
]

\addplot[
    fill={rgb,255:red,72; green,192; blue,170},
    draw=none
] coordinates {(Standalone, 2.95)};

\addplot[
    fill={rgb,255:red,69; green,105; blue,144},
    draw=none
] coordinates {(Network (Float), 0.12)};

\addplot[
    fill={rgb,255:red,239; green,118; blue,122},
    draw=none
] coordinates {(Network (Fixed), 0.06)};

\end{axis}
\end{tikzpicture}
\caption{LEO orbit determination error under different processing strategies (meter).}
\label{fig:orbit_error}
\end{figure}
\begin{figure}[t]
\centering
\begin{tikzpicture}
\begin{axis}[
    ybar,
    bar width=22pt,
    width=10cm,
    height=7.1cm,
    ymin=0,
    ymax=8,
    symbolic x coords={Standalone,Network (Float),Network (Fixed)},
    xtick={Standalone,Network (Float),Network (Fixed)},
    xticklabel style={font=\small},
    ytick=\empty,
    ylabel={},
    y axis line style={draw=none},
    nodes near coords,
    nodes near coords align={vertical},
        nodes near coords style={
        /pgf/number format/fixed,
        /pgf/number format/precision=2
    },
    every node near coord/.append style={font=\small, yshift=3pt},
    axis lines=left,
    axis line style={-},
    tick align=outside,
    enlarge x limits=0.25,
    bar shift=0pt
]

\addplot[
    fill={rgb,255:red,72; green,192; blue,170},
    draw=none
] coordinates {(Standalone, 7.95)};

\addplot[
    fill={rgb,255:red,69; green,105; blue,144},
    draw=none
] coordinates {(Network (Float), 0.21)};

\addplot[
    fill={rgb,255:red,239; green,118; blue,122},
    draw=none
] coordinates {(Network (Fixed), 0.11)};

\end{axis}
\end{tikzpicture}
\caption{LEO time synchronization error under different processing strategies (nanosecond).}
\label{fig:time_error}
\end{figure}
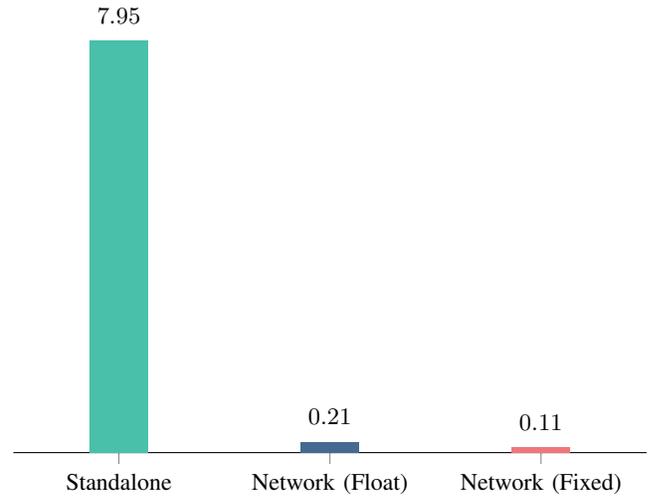

For the algorithm settings, the processing at each epoch is performed over the graph sequence $\{\mathcal{G}^{(t)}\}_{t=1}^{3}$ for at most $K_{\max}=1.2\times10^4$ iterations. Unless otherwise specified, we set the stepsize to $\gamma=0.01$, the momentum parameter to $\theta=0.7$, and the number of consensus rounds to $\mathcal{K}=20$. The GNSS satellite parameters in Algorithm~\ref{alg:gt-atc} are initialized to zero. These values were empirically selected to achieve stable and fast convergence with moderate communication overhead. We report the resulting network-based cooperative solution and compare it against a non-cooperative standalone solution in which each satellite processes its own measurements independently.

\begin{figure}[t]
    \centering
    \includegraphics[width=0.95\linewidth]{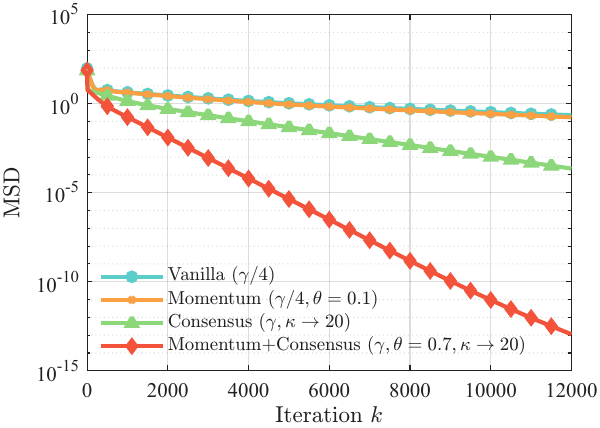}
    \caption{Convergence behavior measured by MSD. We compare the vanilla GT method with its momentum-accelerated variant, a consensus-only variant, and the combined momentum$+$consensus scheme. The latter corresponds to our proposed method, while the others serve as ablations.}
    \label{fig:gt_momentum_convergence}
\end{figure}

\begin{figure}[t]
    \centering
    \includegraphics[width=0.95\linewidth]{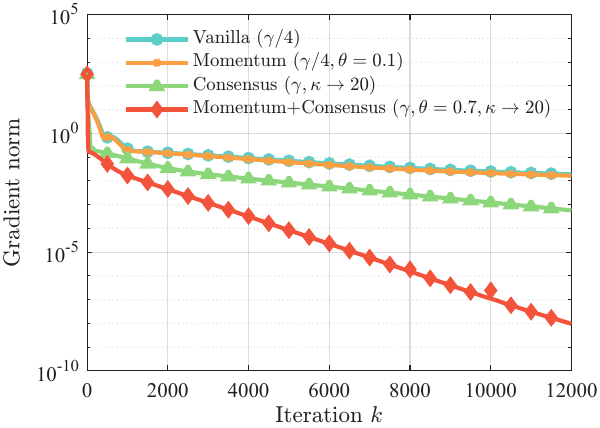}
    \caption{Evolution of the norm of the network-averaged gradient corresponding to the results in Fig.~\ref{fig:gt_momentum_convergence}. We utilize this metric to evaluate the proximity of the solution to first-order optimality.}
    \label{fig:gt_momentum_gradnorm}
\end{figure}

\begin{figure}[t]
    \centering
    \includegraphics[width=0.95\linewidth]{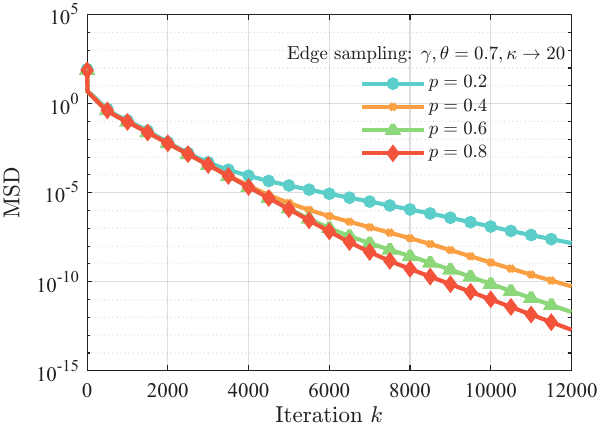}
    \caption{Convergence behavior measured by MSD, for different random link activation probabilities $p\in\{0.2,0.4,0.6,0.8\}$ on $\mathcal{G}^{(t)}$.}
    \label{fig:gt_momentum_convergence_p}
\end{figure}

\begin{figure}[t]
    \centering
    \includegraphics[width=0.95\linewidth]{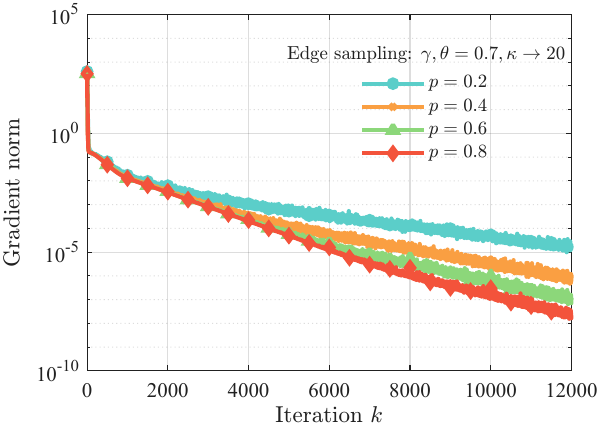}
    \caption{Evolution of the norm of the network-averaged gradient corresponding to the results in Fig.~\ref{fig:gt_momentum_convergence_p}.}
    \label{fig:gt_momentum_gradnorm_p}
\end{figure}

Fig.~\ref{fig:orbit_error} and Fig.~\ref{fig:time_error} compare the orbit determination and time synchronization performance under three processing strategies: standalone per-receiver processing, network-based processing with float ambiguities, and network-based processing with integer-fixed ambiguities. As shown in Fig.~\ref{fig:orbit_error}, standalone processing yields a LEO orbit determination error of $2.95$~m, reflecting the limited accuracy of individual receivers. By jointly estimating shared GNSS satellite clock states across the constellation, the network-based float solution significantly improves the orbit accuracy to $0.12$~m. When integer ambiguity resolution is enabled using the LAMBDA method~\cite{teunnissen1995least}, the network-fixed solution further reduces the orbit error to $0.06$~m, achieving centimeter-level accuracy. Fig.~\ref{fig:time_error} shows the corresponding time synchronization performance. The standalone solution exhibits a timing error of $7.95$~ns. For the network-based solutions, the reported timing errors correspond to the network-consistent clock states. Network-based float processing achieves a timing accuracy of approximately $0.21$~ns, which is further improved to $0.11$~ns after integer ambiguity fixing. Overall, the results demonstrate that network-based onboard GNSS processing enables substantial gains in both orbit determination and time synchronization, with integer ambiguity resolution providing further performance improvements.

We characterize the convergence behavior of our algorithm in Fig.~\ref{fig:gt_momentum_convergence} and Fig.~\ref{fig:gt_momentum_gradnorm}. The former illustrates the mean-square deviation (MSD) with respect to the centralized solution, while the latter depicts the proximity to first-order optimality (gradient norm). Through an ablation study, we demonstrate that our method converges linearly to the centralized solution with significant acceleration. Crucially, this acceleration stems from the synergy between momentum and multi-round consensus; neither component is sufficient on its own. Specifically, the vanilla GT algorithm is restricted to a quarter of the stepsize due to the sparsely connected underlying graph of the LEO constellations. In this regime, incorporating momentum yields only marginal improvements, as stability constraints force the momentum parameter $\theta$ to remain small. However, employing $\mathcal{K}$ consensus rounds reduces the disagreement error between states, enabling the use of a full stepsize $\gamma$. This improved stability accommodates a larger momentum parameter, which significantly accelerates convergence. These results confirm that our algorithm achieves the desired estimation accuracy with reduced overhead for a given tolerance.

In practical LEO constellations, intermittent connectivity, packet loss, and varying link quality may prevent successful information exchange between neighboring satellites. To examine the effect of unavailable exchanges, we further evaluate robustness to intermittent link availability and communication-resource constraints by randomly activating a fraction $p$ of the available inter-satellite links in $\mathcal{G}^{(t)}$ on average during each consensus round. As shown in Fig.~\ref{fig:gt_momentum_convergence_p} and Fig.~\ref{fig:gt_momentum_gradnorm_p}, for all tested activation probabilities $p\in\{0.2,0.4,0.6,0.8\}$, both the MSD and gradient norm decrease steadily toward the centralized solution. Compared with the smoother MSD trajectories, the gradient-norm curves display more pronounced fluctuations due to the randomly varying communication topology. Nevertheless, the proposed method remains stable and convergent even when only $20\%$ of the links are active. Increasing $p$ accelerates convergence and reduces the error attained within a fixed iteration budget. These results indicate that the proposed method maintains stable convergence under stochastic link failures and constrained inter-satellite communication.

Overall, we show that the proposed method
provides a viable and scalable algorithmic framework for
GNSS network processing onboard LEO satellites, where centralized processing and communication resources are limited.

\section{Conclusion}
\label{sec:conclusion}
This paper studied decentralized GNSS network processing for large-scale LEO constellations by modeling the constellation as a cooperative spaceborne network interconnected through inter-satellite links. By exploiting the network structure and the redundancy of GNSS observations across satellites, a network-driven estimation framework was established for joint orbit determination and clock synchronization without centralized data fusion while preserving global consistency. Numerical results demonstrated that cooperative network processing substantially improves both orbit and clock estimation accuracy compared to standalone processing. Within this framework, a momentum-accelerated gradient tracking method enables fully decentralized onboard estimation over a dynamic network and converges rapidly to the centralized benchmark solution. Overall, the proposed framework supports scalable and network-driven GNSS processing for large-scale LEO constellations, enabling high-precision onboard navigation through cooperative estimation in space.

\section*{Acknowledgements}

The authors would like to thank Prof. Peter J. G. Teunissen for conceptual suggestions and valuable discussions on distributed GNSS processing, which helped shape the initial direction of this work.

\appendices
\section{Submatrix Definitions of the Design Matrix}
\label{appendix:A}

\setcounter{equation}{0}
\setcounter{subsection}{0}
\renewcommand{\theequation}{A.\arabic{equation}}
\renewcommand{\thesubsection}{A.\arabic{subsection}}

This appendix provides the block-wise definitions of the submatrices that together form the overall design matrix $\mathbf{A}$.

\noindent \textbf{$\bullet$ LEO orbit–related design matrix:}
\begin{equation}
\mathbf{A}_{\mathrm{leo\text{-}orb}}
=
\operatorname{blkdiag}\limits_{l=1}^{L}
\big(\mathbf{H}_l^{\mathrm{leo\text{-}orb}}\big),
\end{equation}
where
\begin{equation}
\mathbf{H}_{l}^{\mathrm{leo\text{-}orb}} =
\begin{bmatrix}
\mathbf{H}_{l,1}^{\mathrm{leo\text{-}orb}}\\
\vdots \\
\mathbf{H}_{l,F}^{\mathrm{leo\text{-}orb}}
\end{bmatrix}, 
\mathbf{H}_{l,f}^{\mathrm{leo\text{-}orb}} =
\begin{bmatrix}
\mathbf{U}_l & \mathbf{0} \\
\mathbf{U}_l & \mathbf{0} \\
\mathbf{0} & -\dfrac{1}{\lambda_f}\mathbf{U}_l
\end{bmatrix}.
\end{equation}
Here, $\mathbf{U}_l \in \mathbb{R}^{G_l \times 3}$ is the LOS direction matrix for receiver~$l$, with each row representing the LOS unit vector to a visible GNSS satellite.

\noindent \textbf{$\bullet$ LEO clock and hardware–related design matrix:}
\begin{equation}
\mathbf{A}_{\mathrm{leo\text{-}clk}}
=
\operatorname{blkdiag}\limits_{l=1}^{L}
\big(\mathbf{H}_l^{\mathrm{leo\text{-}clk}}\big),
\end{equation}
with
\begin{equation}
\resizebox{.99\hsize}{!}{$
\mathbf{H}_{l}^{\mathrm{leo\text{-}clk}}\! =
\!\! \begin{bmatrix}
\mathbf{H}_{l,1}^{\mathrm{leo\text{-}clk}}\mathbf{C}_{1}^{\mathrm{freq}}\\
\vdots \\
\mathbf{H}_{l,F}^{\mathrm{leo\text{-}clk}}\mathbf{C}_{F}^{\mathrm{freq}}
\end{bmatrix},
\mathbf{H}_{l,f}^{\mathrm{leo\text{-}clk}} \!= \!\! \begin{bmatrix}
1 & 0  &\lambda_f &0\\
1 & 0  &0 &1\\
0 & -\dfrac{1}{\lambda_f}  &0 &0
\end{bmatrix} \otimes \mathbf{1}_{G_l},
$}
\end{equation}
where $\mathbf{C}_{f}^{\mathrm{freq}}$ selects, from $\mathbf{x}_{\mathrm{leo\text{-}clk}}^{(l)}$, the receiver clock and hardware-bias states associated with frequency~$f$, and $\otimes$ denotes the Kronecker product.

\noindent \textbf{$\bullet$ GNSS clock and hardware–related design matrix:}
\begin{equation}
\mathbf{A}_{\mathrm{gnss\text{-}clk}}
=
- \operatorname{col}\limits_{l=1}^{L}
\big(
\mathbf{H}_l^{\mathrm{gnss\text{-}clk}}
\big),
\end{equation}
where
\begin{equation}
\mathbf{H}_l^{\mathrm{gnss\text{-}clk}}
=
\operatorname{col}\big\{
\mathbf{H}_{l,1}^{\mathrm{gnss\text{-}clk}},\ldots,
\mathbf{H}_{l,F}^{\mathrm{gnss\text{-}clk}}
\big\},
\end{equation}
with
\begin{equation}
\mathbf{H}_{l,f}^{\mathrm{gnss\text{-}clk}}
=
\operatorname{col}\big\{
\mathbf{S}_{l,f}^{\phi},\;
\mathbf{S}_{l,f}^{\rho},\;
\mathbf{S}_{l,f}^{d}
\big\},
\end{equation}
\begin{equation}
\mathbf{S}_{l,f}^{\phi}
=
\operatorname{col}_{g\in\mathcal{G}_l}
\big(\mathbf{e}_g^{\top} \otimes
\mathbf{s}_{l,f}^{\phi,g}
\big),
\end{equation}
\begin{equation}
\mathbf{S}_{l,f}^{\rho}
=
\operatorname{col}_{g\in\mathcal{G}_l}
\big(\mathbf{e}_g^{\top} \otimes
\mathbf{s}_{l,f}^{\rho,g}
\big),
\end{equation}
\begin{equation}
\mathbf{S}_{l,f}^{d}
=
\operatorname{col}_{g\in\mathcal{G}_l}
\big(\mathbf{e}_g^{\top} \otimes
\mathbf{s}_{l,f}^{d,g}
\big),
\end{equation}
\begin{equation}
\mathbf{s}_{l, f}^{\phi,g}=\left[1, \quad 0, \quad \lambda_f \mathbf{e}_f^{\top}, \quad \mathbf{0}_{1 \times F}\right],
\end{equation}
\begin{equation}
\mathbf{s}_{l, f}^{\rho,g}=\left[\begin{array}{llll}
1, & 0, & \mathbf{0}_{1 \times F}, & \mathbf{e}_f^{\top}
\end{array}\right],
\end{equation}
\begin{equation}
\mathbf{s}_{l, f}^{d,g}=\left[0, \quad -\frac{1}{\lambda_f}, \quad \mathbf{0}_{1 \times F}, \quad \mathbf{0}_{1 \times F}\right].
\end{equation}
Here, $\mathbf{e}_f \in \mathbb{R}^F$ denotes the $f$-th canonical basis vector, i.e., a vector with a 1 in position $f$ and zeros elsewhere.

\noindent \textbf{$\bullet$ Ionosphere-related design matrix:}
\begin{equation}
\mathbf{A}_{\mathrm{ion}}=\mathrm{blkdiag}_{l=1}^L\left(\mathbf{H}_l^{\mathrm{ion}}\right)
\end{equation}
where
\begin{equation}
\mathbf{H}_l^{\mathrm{ion}}=\left[\begin{array}{c}
\mathbf{H}_{l, 1}^{\mathrm{ion}} \\
\vdots \\
\mathbf{H}_{l, F}^{\mathrm{ion}}
\end{array}\right], \quad \mathbf{H}_{l, f}^{\mathrm{ion}}=
\left[\begin{array}{c}
-\mu_f \\
+\mu_f \\
0
\end{array}\right]
\otimes
\mathbf{I}_{G_l}.
\end{equation}

\noindent \textbf{$\bullet$ Integer ambiguity–related design matrix:}
\begin{equation}
\mathbf{A}_{\mathrm{amb}}=\mathrm{blkdiag}_{l=1}^L\left(\mathbf{H}_l^{\mathrm{amb}}\right),
\end{equation}
with
\begin{equation}
\mathbf{H}_l^{\mathrm{amb}}= \! \mathrm{blkdiag}_{f=1}^F \! \left(\mathbf{H}_{l, f}^{\mathrm{amb}} \right),
\;\;
\mathbf{H}_{l, f}^{\mathrm{amb}} = \! \left[\!\! \begin{array}{c}
\lambda_f \\
0 \\
0
\end{array} \!\!\! \right] \!\! \otimes \! \mathbf{I}_{G_l}.
\end{equation}

\section{Matrices Used for Rank-Deficiency Elimination}
\label{appendix:B}
\setcounter{equation}{0}
\setcounter{subsection}{0}
\renewcommand{\theequation}{B.\arabic{equation}}
\renewcommand{\thesubsection}{B.\arabic{subsection}}

This appendix presents the matrices that underpin the $\mathcal{S}$-system transformation. For convenience, we summarize several notations that will be used throughout this appendix: $\mathbf{c}_n = \left[1, 0, \cdots,0 \right]^\mathrm{T}$, $\mathbf{1}_n = \left[1, 1, \cdots,1 \right]^\mathrm{T}$, $\mathbf{D}_n = \left[\begin{array}{cc}-\mathbf{1}_n &\mathbf{I}_{n-1} \end{array}\right]^\mathrm{T}$, $\mathbf{C}_n = \left[\begin{array}{cc} \mathbf{0}_{n-1} & \mathbf{I}_{n-1} \end{array}\right]^\mathrm{T}$, $\mathbf{F}_n = \left[ \begin{array}{cc} \mathbf{0}_{(n-2) \times 2} & \mathbf{I}_{n-2} \end{array}    \right]^\mathrm{T}$, $\mathbf{E}_n=\mathbf{C}_n \mathbf{D}_n^\mathrm{T}=\mathbf{I}_n-\mathbf{1}_n \mathbf{c}_n^\mathrm{T}$, $\pmb{\mu} =\left[\mu_1, \cdots, \mu_f \right]^\mathrm{T}$, $\bm{\mu}_{\mathrm{IF}}=\frac{1}{\mu_2-\mu_1}\left[\mu_2,-\mu_1, 0, \ldots, 0\right]^\mathrm{T}$, $\bm{\mu}_{\mathrm{GF}}=\frac{1}{\mu_2-\mu_1}[-1,1,0, \ldots, 0]^\mathrm{T}$, $\pmb{\lambda} =\left[\lambda_1, \cdots, \lambda_f \right]^\mathrm{T}$, and $\pmb{\Lambda} =\mathrm{diag}   \left\{\lambda_1, \cdots, \lambda_F  \right\}$. 

The matrices employed to eliminate the rank deficiencies are defined in \eref{eq:V_net} to \eref{eq:S_net_system}. The ambiguity-related blocks are constructed based on the receiver--satellite visibility sets and the adopted pivot selection, and therefore do not admit a universal closed-form expression; their role is to eliminate the ambiguity-related rank deficiencies under the chosen $\mathbf{S}$-basis. For compactness, we omit the explicit listing of these graph-dependent blocks and instead summarize their role in Table~\ref{tab:Rank_def}/Table~\ref{tab:EstPar}.

\begin{figure*}[hbp]
\centering
\begin{equation}
\label{eq:V_net}
\resizebox{.94\hsize}{!}{$
\mathbf{V}
=
\left[
\begin{array}{c|c|c|c|c|c|c}
\mathbf{0} & \mathbf{0}
&\mathbf{0} & \mathbf{0} &\mathbf{0} &\mathbf{0} &\mathbf{0}
  \\[1mm]

\mathbf{1}_L  \otimes \mathbf{I}_{2F+2} &
\mathbf{C}_L \otimes \begin{bmatrix} -1 \\ 0 \\ \Lambda^{-1}\mathbf{1}_F \\  \mathbf{1}_F \end{bmatrix}  &
\mathbf{0} &
 \mathbf{C}_L \otimes \begin{bmatrix} 0 \\ 0 \\\Lambda^{-1} \bm\mu \\ -\bm\mu \end{bmatrix} &
\mathbf{0} &
 -\mathbf{C}_L \otimes \begin{bmatrix} 0 \\ 0 \\\mathbf{I}_F \\ \mathbf{0} \end{bmatrix}  &
\mathbf{0}\\[1mm]

\mathbf{1}_G \otimes \mathbf{I}_{2F+2}  &\mathbf{0} & \mathbf{I}_G \otimes \begin{bmatrix} -1 \\ 0 \\ \Lambda^{-1}\mathbf{1}_F \\ \mathbf{1}_F \end{bmatrix} &\mathbf{0} & \mathbf{I}_G \otimes \begin{bmatrix} 0 \\ 0 \\ -\Lambda^{-1}\bm\mu \\ \bm\mu \end{bmatrix}  &\mathbf{0}  &\mathbf{I}_G \otimes \begin{bmatrix} 0 \\ 0 \\ \mathbf{I}_F \\ \mathbf{0} \end{bmatrix} \\[1mm]

\mathbf{0} &\mathbf{0} &\mathbf{0} &
\left[\begin{array}{cccc}
\mathbf{0}  & \mathbf{0} & \cdots & \mathbf{0}\\
\mathbf{1}_{G_2} & \mathbf{0} & \cdots & \mathbf{0} \\
\mathbf{0} & \mathbf{1}_{G_3} & \cdots & \mathbf{0} \\
& & \vdots & \\
& & & \\
\mathbf{0} & \mathbf{0} &\cdots & \mathbf{1}_{G_L}
\end{array}\right] &\left[\begin{array}{c}
\mathbf{I}_G\left(\mathcal{G}_1,:\right) \\
\mathbf{I}_G\left(\mathcal{G}_2,:\right) \\
\vdots \\
\mathbf{I}_G\left(\mathcal{G}_L,:\right)
\end{array}\right] &
\mathbf{0} &\mathbf{0}\\[1mm]

\mathbf{0} &\mathbf{0} &\mathbf{0} &\mathbf{0}  &\mathbf{0} &
\left[\begin{array}{cccc}
\mathbf{0}  & \mathbf{0} & \cdots & \mathbf{0}\\
\mathbf{I}_F \otimes \mathbf{1}_{G_2} & \mathbf{0} & \cdots & \mathbf{0} \\
\mathbf{0} & \mathbf{I}_F \otimes\mathbf{1}_{G_3} & \cdots & \mathbf{0} \\
& & \vdots & \\
& & & \\
\mathbf{0} & \mathbf{0} &\cdots & \mathbf{I}_F \otimes \mathbf{1}_{G_L}
\end{array}\right] &
\left[\begin{array}{c}
\mathbf{I}_F \otimes \mathbf{I}_G\left(\mathcal{G}_1,:\right) \\
\mathbf{I}_F \otimes \mathbf{I}_G\left(\mathcal{G}_2,:\right) \\
\vdots \\
\mathbf{I}_F \otimes \mathbf{I}_G\left(\mathcal{G}_L,:\right)
\end{array}\right] 
\end{array}
\right]
$}
\end{equation}
\end{figure*}

\begin{figure*}[hbp]
\begin{equation}
\label{eq:S_net_P}
\left(\mathbf{S}^{\perp}\right)^\mathrm{T} = 
\left[
\begin{array}{c|c|c|c|c}

\mathbf{0} & \mathbf{c}_L^{\mathrm{T}} \otimes \mathbf{I}_{2F+2}    &\mathbf{0} &\mathbf{0} &\mathbf{0} \\

\mathbf{0} & \mathbf{D}_L^{\mathrm{T}}  \otimes \begin{pmatrix} \mathbf{0}_{1 \times(F+2)} &\bm{\mu}_{\mathrm{IF}}^{\mathrm{T}} \end{pmatrix}  &\mathbf{0} &\mathbf{0} &\mathbf{0} \\

\mathbf{0} & -\left(\mathbf{c}_L^{\mathrm{T}} \otimes \mathbf{1}_G\right) \otimes \begin{pmatrix}  \mathbf{0}_{1 \times(F+2)} &\bm{\mu}_{\mathrm{IF}}^{\mathrm{T}} \end{pmatrix} & \mathbf{I}_G  \otimes \begin{pmatrix}  \mathbf{0}_{1 \times(F+2)} &\bm{\mu}_{\mathrm{IF}}^{\mathrm{T}} \end{pmatrix}  &\mathbf{0} &\mathbf{0} \\

\mathbf{0} & -\mathbf{D}_L^{\mathrm{T}}  \otimes \begin{pmatrix} \mathbf{0}_{1 \times(F+2)} &\bm{\mu}_{\mathrm{GF}}^{\mathrm{T}} \end{pmatrix}  &\mathbf{0} &\mathbf{0} &\mathbf{0} \\

\mathbf{0} & -\left(\mathbf{c}_L^{\mathrm{T}} \otimes \mathbf{1}_G\right) \otimes \begin{pmatrix}  \mathbf{0}_{1 \times(F+2)} &\bm{\mu}_{\mathrm{GF}}^{\mathrm{T}} \end{pmatrix} & \mathbf{I}_G  \otimes \begin{pmatrix}  \mathbf{0}_{1 \times(F+2)} &\bm{\mu}_{\mathrm{GF}}^{\mathrm{T}} \end{pmatrix}  &\mathbf{0} &\mathbf{0} \\

\mathbf{0} & \mathbf{0} & \mathbf{0} & \mathbf{0} &\mathbf{K}_{\mathrm{rec\text{-}amb}} 
\\

\mathbf{0}  & \mathbf{0} & \mathbf{0} & \mathbf{0} & \mathbf{K}_{\mathrm{sat\text{-}amb}}
\end{array}
\right]
\end{equation}
\end{figure*}

\begin{figure*}[htbp]
\begin{equation}
\label{eq:S_net}
\mathbf{S}
=
\left[
\begin{array}{c|c|c|c|c}
\mathbf{I}_{6L} 
& \mathbf{0} 
& \mathbf{0} 
& \mathbf{0} 
& \mathbf{0}
\\[2mm]

\mathbf{0} 
& 
 \mathbf{C}_L \otimes \left(\begin{array}{cccc}
1 &0 & 0 & 0\\
0 &1 & 0 & 0\\
0& 0& \mathbf{I}_F & 0 \\
0& 0& 0 & \mathbf{F}_F
\end{array}\right) 
& \mathbf{0}
& \mathbf{0}
& \mathbf{0}
\\[3mm]

\mathbf{0} 
& \mathbf{0} 
& \mathbf{I}_G \otimes \left(\begin{array}{cccc}
1 &0 & 0 & 0\\
0 &1 & 0 & 0\\
0& 0& \mathbf{I}_F & 0 \\
0& 0& 0 & \mathbf{F}_F
\end{array}\right)  
& \mathbf{0} 
& \mathbf{0}
\\[3mm]

\mathbf{0} 
& \mathbf{0} 
& \mathbf{0} 
& \mathbf{I}_{\sum_l G_l} 
& \mathbf{0}
\\[3mm]

\mathbf{0} 
& \mathbf{0} 
& \mathbf{0} 
& \mathbf{0} 
& \mathbf{S}_{\mathrm{amb}}
\end{array}
\right]
\end{equation}
\end{figure*}

\begin{figure*}
\begin{equation}
\label{eq:S_net_system}
\resizebox{1.0\hsize}{!}{
$
\begin{aligned}
&\mathcal{S} = \mathbf{I} - \mathbf{V}{\left[\left(\mathbf{S}^{\perp}\right)^\mathrm{T} \mathbf{V}\right]^{-1}\!\!\left(\mathbf{S}^{\perp}\right)^\mathrm{T}} = \mathbf{I} - \mathbf{V}\left(\mathbf{S}^{\perp}\right)^\mathrm{T}=\\
& 
\left[
\begin{array}{c|c|c|c|c}
\mathbf{I}_L \otimes \mathbf{I}_6  &\mathbf{0} 
&\mathbf{0} &\mathbf{0} &\mathbf{0}\\

\mathbf{0} & \mathbf{I}_L \otimes \mathbf{I}_{2F+2}  - \left(\mathbf{1}_L  \mathbf{c}_G^{\mathrm{T}} \right) \otimes \mathbf{I}_{2F+2} - \mathbf{E}_{L} \otimes \left(\begin{array}{cccc}
\mathbf{0} &\mathbf{0} & \mathbf{0} & -\bm{\mu}_{\mathrm{IF}}^{\mathrm{T}}\\
\mathbf{0} &\mathbf{0}  &\mathbf{0} &\mathbf{0}  \\
\mathbf{0} &\mathbf{0} &\mathbf{0}  & \boldsymbol{\Lambda}^{-1} \mathbf{1}_F \bm{\mu}_{\mathrm{IF}}^{\mathrm{T}} - \bm{\Lambda}^{-1} \bm{\mu} \bm{\mu}_{\mathrm{GF}}^\mathrm{T}\\
\mathbf{0} &\mathbf{0}  &\mathbf{0} & \mathbf{1}_F \bm{\mu}_{\mathrm{IF}}^{\mathrm{T}}+\bm{\mu} \bm{\mu}_{\mathrm{GF}}^\mathrm{T}
\end{array}\right) &\mathbf{0} & \mathbf{0}  
&\mathbf{T}_{\text {rec-amb }}\\

\mathbf{0} & - \left(\mathbf{1}_G  \mathbf{c}_L^{\mathrm{T}} \right) \otimes \mathbf{I}_{2F+2}  + \left(\mathbf{1}_L  \mathbf{c}_G^{\mathrm{T}} \right) \otimes \left(\begin{array}{cccc}
\mathbf{0} &\mathbf{0} & \mathbf{0} & -\bm{\mu}_{\mathrm{IF}}^{\mathrm{T}}\\
\mathbf{0} &\mathbf{0}  &\mathbf{0} &\mathbf{0}  \\
\mathbf{0} &\mathbf{0} &\mathbf{0}  & \boldsymbol{\Lambda}^{-1} \mathbf{1}_F \bm{\mu}_{\mathrm{IF}}^{\mathrm{T}}- \bm{\Lambda}^{-1} \bm{\mu} \bm{\mu}_{\mathrm{GF}}^\mathrm{T}\\
\mathbf{0} &\mathbf{0}  &\mathbf{0} & \mathbf{1}_F \bm{\mu}_{\mathrm{IF}}^{\mathrm{T}}+\bm{\mu} \bm{\mu}_{\mathrm{GF}}^\mathrm{T}
\end{array}\right)
& \mathbf{I}_G \otimes \mathbf{I}_{2F+2} - \mathbf{I}_{G} \otimes \left(\begin{array}{cccc}
\mathbf{0} &\mathbf{0} & \mathbf{0} & -\bm{\mu}_{\mathrm{IF}}^{\mathrm{T}}\\
\mathbf{0} &\mathbf{0}  &\mathbf{0} &\mathbf{0}  \\
\mathbf{0} &\mathbf{0} &\mathbf{0}  & \boldsymbol{\Lambda}^{-1} \mathbf{1}_F \bm{\mu}_{\mathrm{IF}}^{\mathrm{T}}- \bm{\Lambda}^{-1} \bm{\mu} \bm{\mu}_{\mathrm{GF}}^\mathrm{T}\\
\mathbf{0} &\mathbf{0}  &\mathbf{0} & \mathbf{1}_F \bm{\mu}_{\mathrm{IF}}^{\mathrm{T}} + \bm{\mu} \bm{\mu}_{\mathrm{GF}}^\mathrm{T}
\end{array}\right)
&\mathbf{0}  & \mathbf{T}_{\text {sat-amb }}
\\

\mathbf{0} 
& \left[\begin{array}{cccc}
\mathbf{1}_{G_1} & \mathbf{0} & \cdots & \mathbf{0} \\
\mathbf{0} & \mathbf{1}_{G_2} & \cdots & \mathbf{0} \\
& & \vdots & \\
& & & \\
\mathbf{0} & \mathbf{0} &\cdots & \mathbf{1}_{G_L}
\end{array}\right]  \otimes \begin{pmatrix} \mathbf{0}_{1 \times(F+2)} &\bm{\mu}_{\mathrm{GF}}^{\mathrm{T}} \end{pmatrix}  
& -\left[\begin{array}{c}
\mathbf{I}_G\left(\mathcal{G}_1,:\right) \\
\mathbf{I}_G\left(\mathcal{G}_2,:\right) \\
\vdots \\
\mathbf{I}_G\left(\mathcal{G}_L,:\right)
\end{array}\right]  \otimes \begin{pmatrix} \mathbf{0}_{1 \times(F+2)} &\bm{\mu}_{\mathrm{GF}}^{\mathrm{T}} \end{pmatrix} 
& \mathbf{I}_{\sum_l G_l} 
& \mathbf{0}
\\

\mathbf{0} & \mathbf{0}  & \mathbf{0}  & \mathbf{0} & \mathbf{T}_{\text {amb }}
\end{array}
\right]
\end{aligned}
$}
\end{equation}
\end{figure*}

{
\bibliographystyle{IEEEtran}
\bibliography{IEEEfull}
}

\end{document}